# SAMPLE-BASED SUBSAMPLING STRATEGIES TO IDENTIFY MICROPLASTICS IN THE PRESENCE OF A HIGH NUMBER OF PARTICLES USING QUANTUM-CASCADE LASER-BASED INFRARED IMAGING

Adrián López-Rosales, José M. Andrade*, Borja Ferreiro, Soledad Muniategui

Group of Applied Analytical Chemistry, Institute of Environmental Sciences (IUMA), University of A Coruña, Campus da Zapateira s/n, E-15071 A Coruña, Spain

(*) andrade@udc.es, fax: +34981167065

## ABSTRACT

Microplastics (MPs) are ubiquitous in all ecosystems, affecting wildlife and, ultimately, human health. The complexity of natural samples plus the unspecificity of their treatments to isolate polymers renders the characterization of thousands of particles impractical for environmental monitoring using conventional spectroscopic techniques. Two primary solutions are to analyze a small fraction of the sample or to measure only a subset of particles present over a holder, known as subsampling.

A strategy to subsample reflective Kevley slides and gold-coated filters using quantum-cascade laser-based infrared imaging is proposed here, as this technology is a promising tool for MPs monitoring. In contrast to most previous approaches that struggle to propose general subsampling schemes, we introduce the concept of sample-based subsampling. This can be applied *ex-ante* always and it highlights the best subsampling areas for a sample after a preliminary assay to count the total number of particles on a holder. The error at this stage acts as a proxy to minimize errors when evaluating the number of particles and MPs, significantly enhancing the feasibility of large-scale MPs monitoring. The predictive ability of the approach was tested for fibres and fragments, for total amounts of particles and MPs. Further, the evaluations were disaggregated by size and polymer type. In most situations the reference values were contained in the confidence intervals of the predicted values (often within the 68 % ones) and relative errors were lower than 25 %. Exceptions occurred when very scarce (one or two) items of a given size or polymer were present on the overall holder. The approach was compared to other systematic *ad-hoc* strategies.

## KEY WORDS

Microplastics; Subsampling; reflective slide; gold-coated filter; LDIR; subsampling uncertainty.

**GRAPHICAL ABSTRACT**

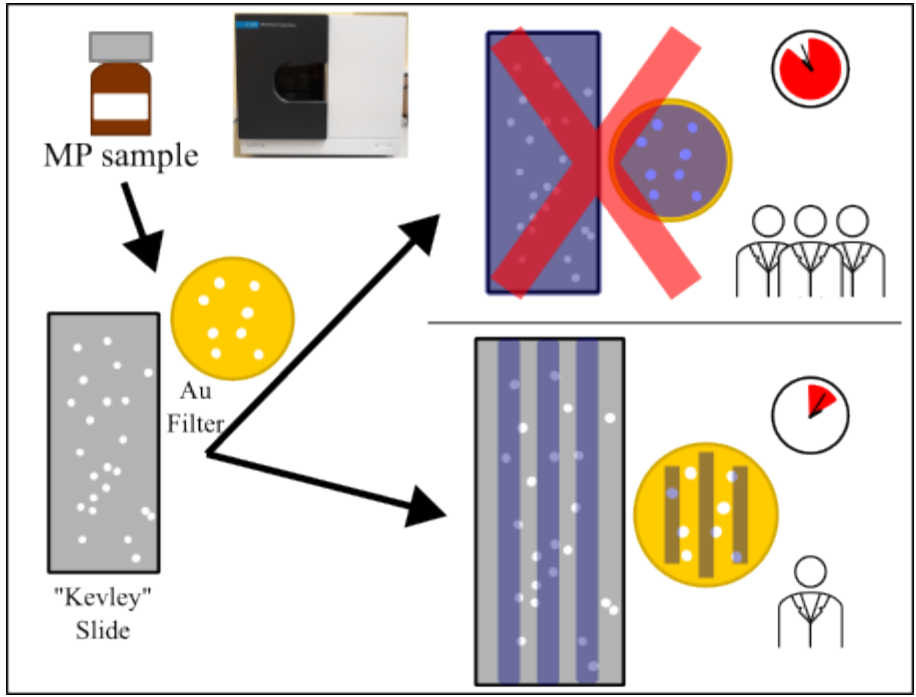

# 1. INTRODUCTION

Microplastics (MPs) pollution poses a significant environmental risk that affects ecosystems worldwide. Those tiny plastic particles, less than 5 mm in diameter, are found in oceans, rivers, soils, and even in the air we breathe. Their pervasive presence may threat wildlife and human health, as they can be ingested, along with many other pollutants adsorbed on them, leading to potential health risks [1]. Global plastic production soared from approximately 270 million metric tons (Mt) in 2010 to 413.8 Mts by 2024 [2]. It was estimated that between 1950 and 2015 around 6,300 Mt of plastic were discarded as waste and that about 12.000 Mt are expected by 2050 [3].

Following, monitoring MPs in different environmental compartments is becoming a must to study their abundance and typology and to assess potential animal/human health hazards. However, the complexity of many natural samples like, e.g., atmospheric deposition and sediments or soils, where the MPs are accompanied by huge amounts of matrix particles -both organic and inorganic-, makes sometimes impractical analyzing the very many particles that appear in the final measurement platform (being it a filter or a reflective slide). For instance, up to 30,000 particles/filter may easily have to be inspected and their characterization would last for a week irrespectively of the spectroscopic technique under consideration, most commonly Raman or FTIR (and regardless of using single-spot MCT, linear arrays or FPA detectors) [4].

When the new technology termed quantum-cascade laser-based infrared transflectance (or reflectance-absorbance) imaging (QCL-IRRAS), or LDIR (coined as its commercial name, laser direct infrared), is used the time required for the measurement decreases but even several days will be needed. Accordingly, monitoring –a major objective of governmental environmental assessments- cannot be carried out due to the huge amount of time required to characterize high numbers of particles in many samples, collected at different locations and several sampling campaigns. To ameliorate this problem, some international approaches proposed cost-effective and fast working protocols [5] but there are still many issues to be solved, even for the "simple" approach they consider. REMARCO (the Research Network of Marine-Coastal stressors in Latin America and the Caribbean, in close collaboration with the International Agency for the Atomic Energy – Environmental laboratories, ARCAL program-) is also developing standardized protocols than would be applicable at many countries (current efforts can be consulted at [6]).

In addition, a current move nowadays consists of measuring as smallest particles as possible to get insights into toxicological issues. This poses further practical complications because the number of particles increases exponentially following a power function (and so does the time required for their characterization) as particle size decreases [7], if that is finally possible at all due to time, computation and software constraints (e.g., the limited capability to stich altogether a high number of small high-magnified visual images to get the final overview of a filter). Possible solutions are not straightforward and two major options are currently available:

i) To analyze only a small fraction of the processed sample. However, this involves a serious risk when the final result is extrapolated [4]. This is particularly problematic whenever few particles of a given size –usually relatively large- are to be estimated in a sample [4,8]. There are also problems to assure representativeness of the aliquot withdrawn for measurement [9]. Further, aliquoting is not recommended by the EU unless strictly needed [10]. In particular, when chemical imaging is performed using a LDIR system, samples are deposited over a reflective slide and the use of aliquots has been common [11–15], with associated representativity problems. A possible solution to avoid aliquoting is to perform a mild evaporation of the final suspension and deploy the final volume on a reflective slide [16]. Despite extrapolation is avoided using this approach and representativeness is granted (as no aliquot is used) the method is a bit slow and it does not solve the need for measuring up to thousands of particles.

ii) To measure only a subset of the particles deployed in the filter or reflective slide. This has been termed "subsampling" or "subsampling during analysis" [4] and it is the approach that will be studied here. Therefore, in the following it will be considered that all particles from the final treatment of a sample were transfered to a proper flat, reflective platform. The two approaches available nowadays for LDIR systems will be considered: reflective microscopy Kevley slides and gold-coated polymeric (PET) filters.

Several subsampling strategies were tested by Thaysen *et al*. [17] when using filters and it was found that homogeneity in the distribution of particles over the filters can rarely be assured and, therefore, that some strategies are prone to fail. Random subsampling of particles is advisable [8,18] but in many available IR measuring systems this is not possible due to software or mechanical limitations. Thus, a measurement strategy based on random windows to avoid bias was proposed by Schwaferts et al. [8] with good results; although with the operational problem that a high number of particles need to be measured to get representativeness and, so, lengthy measurements are required. Brandt *et al*. [19] made a comprehensive comparison between systematically -and randomly- deployed areas and found that none or the tested approaches outperformed each other, but for specific edge scenarios. They recommended measuring at least 7,000 particles (or 50 % of the particles for the random models), or 50 % of the filter area to get relative errors below 20 %. This requirements might still impose serious limitations in the overall working speed, as measuring 50 % of the filter area may be sometimes too time-consuming if many particles and samples are to be processed, up to days [20]. They also found that the standard deviation of subsampling increased significantly with fewer measured areas. As a compromise, at least 20 % of the filter area would be subsampled to account for the 5-50 µm particle size fraction (and measure at least 1,000 particles throughout the overall filter area). They argued that for particle sizes >50 µm the overall filter might have to be measured. These conclusions differ from a previous study focused on sediments reporting that 12.5 % fractions of the filter were the best trade-off between time, workload and reliability [20]. However, they acknowledged that it was not the most accurate option as for some samples and polymers they got >50 % errors, a reason being that they used wedge ("cheese portions") measurement areas. These had been suggested previously [4] although they are usually not feasible due to current software constraints. Huppersstberg and Knepper [21] suggested helical patterns of small squares to reduce grouping and avoid segregation errors. This pattern was adapted slightly to address some practical applications [18,22].

The present work investigates whether subsampling can be applied to accelerate the measurement of MPs in environmental samples using LDIR imaging systems, a relatively new IR technology commercialised by Agilent less than 5 years ago. This new IR system is particularly fast (it can characterize around 8,000-9,000 particles per 24 h) and opens a path for monitoring purposes. In particular, we focus first on the use of rectangular reflective microscopy-type slides where all the particles of the processed sample are deployed. Then, we expand the findings to the use of gold-coated filters.

To the best of our knowledge this issue has not still been addressed systematically despite the community of LDIR users expands rapidly. We found only an LDIR application where 20 random squares were applied [12],with the only comment that a very limited 8 % of the area had been considered. The present work aligns with the conclusions of two relevant studies [4,8] that pointed out that subsampling approaches largely depend on the available instrumentation, and with recent European Union guidelines tailoring working procedures that need to be deployed and, most important, validated [10]. Those guidelines state that, either all the particles in the plate (filter) must be analysed (if possible), or at least 10 % of them (or 20 % of the total area) must be characterized otherwise. Therefore, establishing subsampling approaches for the new LDIR system merits detailed studies, which are presented here.

Contrary to most approaches that struggle to propose general-purpose subsampling schemes, we introduce the concept of sample-based subsampling strategy. Here, the best subsampling approach (i.e., the number of small areas where all particles have to be characterized to get the lowest possible error) is decided on a per sample basis after a preliminary assay to count the total number of particles in that sample. Hence, for the purposes of this work, sample-based subsampling can be defined as an approach in which the regions where an exhaustive search for MPs for a given sample are decided after an initial particle count across the entire sample. This would minimize bias and optimize representativity, getting a tradeoff between analytical efficiency and accuracy.

This had been suggested elsewhere when concluding that any subsampling strategy should be based on empirically determining a pattern that minimizes over- and under-estimations of particles [17], problem until now was how to perform this. Another sample-based proposal had been implicitly depicted when "*on-the-fly*" [*sic*] Raman protocols were discussed [4,8]. Nevertheless, they pointed out the long time required to carry out preliminary evaluations. On the contrary, such an evaluation is feasible indeed using the LDIR system, as this step is usually fast allowing for an estimation of the error that might be associated reasonably to the final extrapolation and, so, developing a sound criterion to decide on the number of regions to characterize.

The goal is to ensure that subsampling is not performed by a unique predefined pattern but it is instead guided by the actual spatial distribution of the particles. Unlike pre-determined systematic subsampling strategies, the sample-based approach dynamically adapts to each sample, accounting for variations in particle distribution.

## 2. EXPERIMENTAL PART

### 2.1. Samples, reagents and materials.

Several types of true environmental samples which usually render thousands of particles in the filters were considered. They bracket a range of different applications and we think they can be used as exemplary case studies. Sample #1 corresponds to a superficial oceanic water sample (collected in the ría of Pontevedra from a small boat, Southern Galicia, Spain); sample #2 was a pool of mussels (*Mityllus Galloprovincialis* from the shoreline of the ría da Coruña, Northwestern Galicia, Spain), and samples #3 and #4 correspond to superficial river waters (Mero and Mendo rivers, giving rise to the Cecebre water reservoir, close to the Metropolitan area of A Coruña, (NW Galicia, Spain). Sample #5 is about gastrointestinal tracts of crayfish (Mero and Mendo rivers) and sample #6 proceeds from atmospheric deposition (collected in a dedicated cabin for atmospheric pollution monitoring, located at the Institute of Environmental Sciences, IUMA, Oleiros, A Coruña, Spain). The procedural blanks were obtained throughout the overall analytical process (enzymatic and oxidative treatment).

Water and bulk atmospheric samples followed a surfactant plus oxidative treatment as described previously [23], whereas mussels and crayfish required an enzymatic-oxidative treatment outlined elsewhere for biota samples [16].

Procedural blanks were included in this study primarily as an example of a sample with very few MPs, allowing us to verify whether the subsampling approach remains effective even under low-particle conditions. Two representative blanks are reported: one corresponding to the enzymatic-oxidative treatment and another to the surfactant-oxidative treatment. These were selected because they reflect the expected background levels for different matrices processed with the same protocol. While procedural blanks inherently serve to monitor potential procedural contamination, their main purpose here is to demonstrate that the subsampling strategy provides reliable results even when particle counts are minimal.

The transfer protocol to Kevley slides was the so-called "Syncore method" described in previous works [16], while the transfer to gold-coated filters involved resuspending the filter cake in 50 % ethanol and filtering it through a gold-coated filter. In this study, a single reflective slide or a single gold-coated filter was used per sample.

The reagents were sodium dodecyl sulfate (SDS ≥ 98.5 % purity), $H_2O_2$ (≥ 30 %), TRIS (tris-(hydroxymethyl)aminomethane), and protease from Streptomyces griseus (Type XIV activity ≥ 3.5 units/mg), all from Sigma-Aldrich. Flat reflective slides (MiRR, Kevley Technologies, Chesterland, USA) and 0.8 µm gold-coated filters (25 mm diameter, from i3 Membrane, Germany) were used. The polymers identified in this work were EVA (Ethylene-vinyl acetate), PA (Polyamide), PE (Polyethylene), PC (Polycarbonate), PET (Poly(ethylene terephthalate)), PMMA (Poly(methyl methacrylate)), POM (Poly(oxymethylene)), PP (Polypropilene), PS (Polystyrene), PU (Polyurethane), PVC (Poly(vinyl chloride)), ABS (Acrylonitrile butadiene styrene), PTFE (Poly(tetrafluoroethylene)), polyacrylate and alkyd varnish.

## 2.2. Instruments

An automatic evaporation system composed of a V-800/805 vacuum controller, a vacuum line and an R-12 analyst Syncore device, plus dedicated glass containers (residual volume 1.0 mL, Büchi, Switzerland); a Rotabit P incubation system (Selecta, Spain), with temperature and agitation controls; a Pobel vacuum filtration system combined with a Millipore vacuum pump (Millipore, Ballerica, MA, model WP6122050); a 3000867 Selecta ultrasonic bath (Barcelona, Spain); and a 2001 pH-meter from Crison (Barcelona, Spain), were employed throughout.

A quantum cascade laser-based infrared 8700 LDIR device from Agilent was used. It is a high-speed monitoring instrument capable of detecting particles as small as 10 µm in size. The measuring conditions were: 1800-975 $cm^{-1}$ spectral range, spectral resolution 8 $cm^{-1}$, one scan per spectrum, sensitivity was set to 5 (v. 1.6, Clarity® software). The automatic counting of particles was made with Clarity software using the particle analysis tool, considering 1499 $cm^{-1}$ as the initial imaging wavenumber. The size of the particles scanned ranged from 5000 to 20 µm. The identification of the polymers was done using built-in databases complemented with in-house spectra of different materials; the match index was set to 0.90 for positive identifications, since this reduces the number of false positives [23].

As it was stated previously [9], the detection of smallest particles is limited by instrumental capabilities, like the size of the pixels of the detectors, shadowing or edge effect that difficults the sharp definition of the borders of the particle, light beam focusing abilities, etc. Although the LDIR system can image a particle down to <10 µm when individual optical magnification and individual spot measurement are selected, these options are not suitable for the automatic analyses of thousands of particles. Therefore, it is likely that the number of particles and their characterization become faulty in the 20-10 µm size range. This is, of course, the concept of the instrumental limit of detection and, therefore, we decided not to consider this size range and focus only on >20 µm particles.

Despite the LDIR being a fast instrument, capable of characterising up to 8,000-9,000 particles/24 h, in many environmental samples it is required to address >30,000 particles deployed on the slide (as we experienced for air deposition and fine seabed sediments).

## 2.3. Quality control

To prevent contamination by external particles, all laboratory materials were washed with 17 % HCl for 24 h [24]. Alternatively, 50 % ethanol (EtOH), 2 % SDS, and a small amount of alkaline soap can be used in an ultrasonic bath for 1 h. All materials were rinsed with prefiltered (0.45 μm) 50 % EtOH, followed by Milli-Q water, both before and during all procedural steps to avoid cross-contamination. The gold-coated and stainless-steel filters were rinsed with 50 %, and sonicated for 10 min in 50 % ethanol before use, respectively. All reagents were filtered through 0.45 μm glass fibre filters [10,18]. Whenever possible, plastic materials were replaced with glass or metal alternatives. All containers and glassware were covered with aluminum foil during storage, use, and while working in fume hoods to prevent airborne contamination. The transfer protocol to Kevley slides and gold-coated filters was carried out in a laminar flow hood (Noxair).

# 3. RESULTS AND DISCUSSION

## 3.1. Sample-based subsampling strategy

Many subsampling strategies would benefit from initial estimates on the number and distribution of the particles (ideally of the MPs) on the filter or supporting plate as this would allow to define some *ad-hoc* criterion to select a given approach (regardless of being a systematic pattern or a random distribution of small areas) and setting some kind of statistical distribution [8,19]. However, the recommended large number of particles (ca. 7,000) or minimum size of sampled filter area (ca. 50 % of the filter) makes this option unpractical in most situations (FTIR-FPA detection, FTIR single-spot detection, Raman, etc.).

Furthermore, it is recognized nowadays that each filter is unique due to the inhomogeneity of the distribution and position of the particles, the nature of the environmental particles, their size, shapes and densities, the remains of the (in)organic matrix, etc. [4,17]; leading to a specific spatial distribution that cannot be anticipated. In some cases, the particles may accumulate at the center of the filter [17], with corresponding risks of particle overlapping, aggregation, masking, etc.

The concept of sample-based subsampling strategy developed here for LDIR devices evaluates, on a per-sample basis, the spatial information that it is gained experimentally from the complete image of the reflective support (hereinafter, "plate") where the particles are deposited. The particles of a give size range are detected and counted automatically, and this number is used to determine the number of subareas that are required to approach that value, given a maximum bias requirement (e.g., < 20 %). This workflow adheres to recommendations given recently [17] although without trying to get a general behaviour for a kind of samples, as we consider each sample as an individual study.

Note that this step only involves detection of the particles, not their characterisation and, so, it does not constitute a bottleneck for the overall process (>20.000 particles can be detected and counted in ca. 30 h). Besides, this will assure that any region and any particle have –initially- the same probability to be selected and, thus, we avoid initial statistical biases [8]. Following, simple calculations allow determining which set of subareas approaches the overall number of particles best for the sample at hand. These are the areas where all the particles have to be characterized to identify the MPs. Hence, the time-consuming stages to characterize spectroscopically each and every particle are carried out only on the selected subareas. This is why this strategy is sample-based and it can fit any (uncontrollable) deposition pattern that may occur in the plate. Other published approaches are not based on the *ad-hoc* sample distribution of the particles.

No doubt this sampling strategy is based on some fundamental assumptions, common to other strategies as well: i) that the topological distribution of the MPs follows that of the other particles in the slide (or filter); in other words, that there is not a sound reason that makes MPs to concentrate/ disperse/deposit (etc.) on areas different from those where the other particles are; (ii) that small and large particles behave in the same way and that there is not segregation by size [8]; and (iii) that the particles do not accumulate in a limited region of the plate and are not superimposed. In our experience, and in most of the papers we have reviewed, this is not likely to happen when reflective slides are used, mostly if the final suspension that results after the sample treatment is poured carefully all over the reflective surface and the usual 50 % ethanol:water solvent is allowed to evaporate gently and the deposition is done sequentially. This corresponds to the requisite that filtering is done as correctly as possible [4]. Therefore, providing this manual transference is done correctly, the sample-based subsampling strategy seems reasonable. To assure good particle transferences the use of either a surfactant (0.05 % w/w) [21] or 96 % ethanol –best suited for evaporation in the reflective plate- when filtering through gold-coated filters were recommended [16].

Additional advantages of the sample-based approach are that it is valid for every size range that may be of interest (as this is fixed before the initial detection of particles is done by the LDIR system) and that one can decide on the overall size of the scanning area given their own limitations of time, workload, etc. since an approximate error will be evaluated for each decision using straightforward calculations (see next section).

### 3.2. Calculations

Once the particles of a given target size range are detected and their total number throughout the overall filter/plate counted by the LDIR, the overall area of the holder is divided in different subregions. For the sake of simplicity, pragmatic workload and measuring times we set 10 parallel adjacent subregions or slices (although this can be tailored by the analyst), each representing 10 % of the surface. The number of particles on each slice is counted and subsequently extrapolated to the overall plate by multiplying times 10. Following, the error in the extrapolated prediction of the number of particles when only a slice (denoted as $i$) is employed becomes $Error_{slice\ i}\ (\%) = 100 \cdot \frac{(10 \cdot estimated\ particles_{slice\ i}) - all\ particles}{all\ particles}$ . It is expected that some slices will not be good enough to approach the number of particles (e.g., a slice very close to a border in a reflective plate will almost always underestimate the values because we usually avoid depositing the particles too close to the border of the plate) and, so, the most suitable ones should be ascertained. Usually (but not always) those at the centre are best. To adhere to most general conventions given in the references above and in order to cover as much area as possible to get statistically sound results [4] we decided to measure at least 40 % of the reflective slide (4 slices); this clearly exceeds the minimum 20 % recommendation of the EU [10], but it is still a reasonable trade-off when working with LDIR. They are selected considering the four slices whose extrapolations yield the lowest errors when evaluating the total number of particles: those two below the median and those two above it. The final results for both the number of particles and total MPs of the sample will be the median of the four extrapolated values. For known samples or those for which all the MPs were determined in advance (like those shown in the present study), the error per slice in the identification of each type of polymer (termed as $k$) can be calculated as

$$Error_{(polymer\ k)} = 100 \cdot \frac{(10 \cdot estimated\ microplastics_{(polymer\ k)}) - total\ microplastics_{(polymer\ k)}}{total\ microplastics_{(polymer\ k)}}$$ , and

$$Recovery_{(polymer\ k)} = 100 \cdot \frac{(10 \cdot estimated\ microplastics_{(polymer\ k)})}{total\ microplastics_{(polymer\ k)}}.$$

It is worth noting that this error can be calculated here because we characterized the MPs throughout overall filters/plate, but this would not be the case for routine, unknown samples.

Please, bear in mind that in previous works the total number of particles was estimated only *ex-post* but no reliable quantity was available *ex-ante*, except for Brand *et al.*, [4] as they also used fully-measured previous filters, but that situation will not be feasible for true samples in routine work. On the contrary, the sample-based approach shown here can be applied *ex-ante* always.

### 3.3. Subsampling in reflective Kevley slides

The use of reflective Kevley slides involves depositing aliquots of a suspension where the particles are over a slide, waiting for solvent evaporation, and repeating the process until the suspension is transferred to the slide, some cleaning of the initial container is recommended with the corresponding transference of solvent to the slide. In our case the suspensions were obtained using the "Syncore method" [16]. This procedure naturally distributes the particles irregularly, asymmetrically and in a way difficult to subsample following predefined schemes (see **Figure 1**). After signaling the ten slices in the computer, the number of particles of a selected size range is counted on each one automatically by the LDIR system and their extrapolated values are calculated manually. The number of slices whose particles need to be characterized is determined by the area that the analyst wants to study and the error (s)he can accept. In this study we consider 40 % of the overall surface (4 slices) as a reasonable guideline, although in some cases this might be reduced, we also would like to have relative errors <20 %.

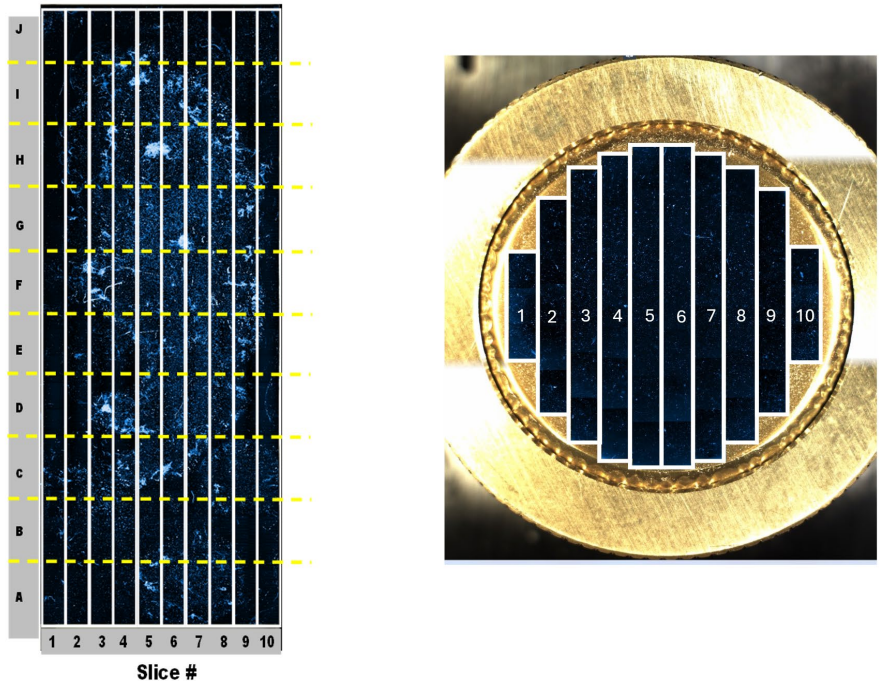

**Figure 1**: Example of the distribution of particles on a reflective microscopy slide and, on a gold-coated filter, along with the 10 slices that are considered to count the number of particles on them.

**Table 1** summarizes the studies performed on four widely different typical environmental samples (as detailed in section 2) and two blanks. The slices were ordered by the number of extrapolated particles and the four with the lowest relative errors, two above and two below the median, were selected. Then, the median of the four predictions and its relative overall error were calculated.

It can be seen that the overall prediction errors (reference – experimental value) for the total number of particles in the samples are excellent, lower or close to 5 %, which outperforms other studies. Note also that if we were to apply the EU guidelines [10] by which only 20 % of the surface would be analyzed, we would even had obtained very satisfactory results (consider two slices, those above and below the median), which is highly encouraging. The overall errors increased for the blanks, as expected because of two reasons: the much lower number of particles and their highly uneven and sparse distribution, which leads to not-so-good estimations even when 40 % of the reflective slide is surveyed. Nevertheless, the values are excellent for most common applications. However, this is not really a major problem because usually we will not perform subsampling when evaluating blanks. The analysis times are usually short and we are indeed interested on characterizing all the few MPs that they contain.

The next step is to estimate the total amount of MPs. Recall that for unknown samples we would determine the MPs on the slices selected according to the strategy given in the paragraphs above (which have a grey background in the tables) and that the "true" values will be unknown.

**Table 2** presents the results for the MPs estimated for the samples and the blanks. They are not as good as for the total number of particles, but they are still very satisfactory because the maximum errors were lower than 22 % (that high value occurred for a sample and a blank). The explanation for the lower performance in blank#2 is the very low number of MPs and their uneven distribution (and recall that blanks would not usually be subsampled). Curiously, the slices selected when the number of particles is taken into account were #4, 7, 8 and 9 but slices 5 and 6 would be better when the MPs are considered (but that information, of course, will not be known *ex-ante* for new samples). The interesting point here is that the sample-based subsampling approach might be quite acceptable even for the extreme situations the blanks represent (mostly if blanks are evaluated considering more than four slices).

The relative standard deviation for the means (RSD) of the predictions using 4 slices are very good when considering the number of particles (<10 %) although higher when considering the MPs (7-44 %). This can be explained because the number of MPs is definitely much smaller than the number of particles (less than 1 % of the particles are MPs) and so it is easier to predict the latter than the former. This is one of the reasons why we suggest using at least four slices (40 % of the total surface).

**Table 1:** Evaluation of the total number of particles in a reflective slide by means of subsampling. The cells corresponding to the selected slices have a grey background. RSD stands for relative standard deviation of the mean, given as a percentage

| Sample 1 | | | Sample 2 | | | Sample 3 | | | Sample 4 | | | Blank 1 | | | Blank 2 | | |
|---|---|---|---|---|---|---|---|---|---|---|---|---|---|---|---|---|---|
| Slice # | particles | % error | Slice # | particles | % error | Slice # | particles | % error | Slice # | particles | % error | Slice # | particles | % error | Slice # | particles | % error |
| 4 | 45480 | 37.7 | 7 | 26940 | 22.2 | 7 | 35040 | 29.2 | 8 | 50070 | 27.7 | 5 | 25340 | 33.3 | 10 | 20790 | 64.6 |
| 3 | 44520 | 34.8 | 9 | 26230 | 19.0 | 10 | 34280 | 26.4 | 5 | 47850 | 22.0 | 3 | 25300 | 33.1 | 2 | 17370 | 37.6 |
| 5 | 40840 | 23.6 | 10 | 25770 | 16.9 | 5 | 34200 | 26.1 | 4 | 46650 | 19.0 | 4 | 22590 | 18.8 | 1 | 15700 | 24.3 |
| *6* | *36310* | *9.9* | *4* | *24380* | *10.6* | *6* | *31200* | *15.0* | *9* | *44670* | *13.9* | *8* | *21360* | *12.3* | *4* | *11700* | *-7.3* |
| *2* | *33930* | *2.7* | *8* | *23670* | *7.4* | *9* | *29320* | *8.1* | *6* | *42730* | *9.0* | *6* | *20580* | *8.2* | *8* | *11420* | *-9.6* |
| *7* | *33370* | *1.0* | *6* | *21670* | *-1.7* | *4* | *26470* | *-2.4* | *3* | *41670* | *6.3* | *1* | *19880* | *4.5* | *9* | *10810* | *-14.4* |
| *8* | *29050* | *-12.1* | *5* | *19110* | *-13.3* | *3* | *25410* | *-6.3* | *2* | *39020* | *-0.5* | *2* | *19740* | *3.8* | *7* | *10520* | *-16.7* |
| 9 | 25120 | -24.0 | 3 | 18860 | -14.5 | 8 | 23430 | -13.6 | 7 | 38490 | -1.8 | 7 | 18240 | -4.1 | 5 | 10180 | -19.4 |
| 1 | 24710 | -25.2 | 2 | 17900 | -18.8 | 2 | 19640 | -27.6 | 10 | 28870 | -26.4 | 9 | 9780 | -48.6 | 6 | 9410 | -25.5 |
| 10 | 16990 | -48.6 | 1 | 15940 | -27.7 | 1 | 12240 | -54.9 | 1 | 12130 | -69.1 | 10 | 7340 | -61.4 | 3 | 8370 | -33.7 |
| Total particles | 33032 | | | 22047 | | | 27123 | | | 39215 | | | 19015 | | | 12627 | |
| Median selected slides | 33650 | | | 22670 | | | 27895 | | | 42200 | | | 20230 | | | 11115 | |
| Recovery median (%) | 101.9 | | | 102.8 | | | 102.8 | | | 107.6 | | | 106.4 | | | 88.0 | |
| Average selected slides | 33165 | | | 22208 | | | 28100 | | | 42020 | | | 20390 | | | 11113 | |
| RSD (%) | 9.1 | | | 10.6 | | | 9.4 | | | 5.7 | | | 3.7 | | | 4.9 | |
| Recovery average | 100.4 | | | 100.7 | | | 103.6 | | | 99.6 | | | 107.2 | | | 88.0 | |
| Error (%) | 0.4 | | | 0.7 | | | 3.6 | | | 0.6 | | | 7.2 | | | -12 | |

**Table 2:** Evaluation of the total number of microplastics in a reflective slide by means of subsampling. The cells corresponding to the selected slices (shown in Table 1) have a grey background. RSD stands for relative standard deviation of the mean, given as a percentage.

| | Sample 1 | | | Sample 2 | | | Sample 3 | | | Sample 4 | | | Blank 1 | | | Blank 2 | |
|---|---|---|---|---|---|---|---|---|---|---|---|---|---|---|---|---|---|
| Slice # | MPs | % error | Slice # | MPs | % error | Slice # | MPs | % error | Slice # | MPs | % error | Slice # | MPs | % error | Slice # | MPs | % error |
| 10 | 360 | -69.6 | *5* | *140* | *-46.6* | *6* | *380* | *-47.7* | 1 | 170 | -67.4 | 7 | 10 | -80 | 10 | 10 | -69.7 |
| 9 | 400 | -66.2 | *6* | *150* | *-42.7* | 1 | 380 | -47.7 | 10 | 370 | -29.1 | 9 | 10 | -80 | 2 | 10 | -69.7 |
| *2* | *550* | *-53.6* | 1 | 220 | -16 | 8 | 570 | -21.5 | 5 | 430 | -17.6 | *6* | *30* | *-40* | 3 | 10 | -69.7 |
| 1 | 650 | -45.1 | 2 | 240 | -8.4 | 7 | 610 | -16 | *3* | *470* | *-10* | *8* | *40* | *-20* | *8* | *20* | *-39.4* |
| *8* | *1180* | *-0.4* | *4* | *260* | *-0.8* | 9 | 760 | 4.7 | *6* | *470* | *-10* | 10 | 40 | -20 | 5 | 30 | -9.1 |
| *6* | *1530* | *29.1* | 3 | 270 | 3.1 | *3* | *790* | *8.8* | *9* | *470* | *-10* | 4 | 50 | 0 | 6 | 30 | -9.1 |
| 5 | 1650 | 39.2 | 10 | 310 | 18.3 | 10 | 810 | 11.6 | *2* | *540* | *3.4* | 5 | 50 | 0 | *9* | *40* | *21.2* |
| *7* | *1700* | *43.5* | 7 | 330 | 26 | 5 | 840 | 15.7 | 7 | 620 | 18.8 | *2* | *60* | *20* | *7* | *40* | *21.2* |
| 3 | 1850 | 56.1 | *8* | *350* | *33.6* | 2 | 1010 | 39.1 | 4 | 700 | 34.1 | *1* | *80* | *60* | *4* | *60* | *81.8* |
| 4 | 1980 | 67.1 | 9 | 350 | 33.6 | *4* | *1110* | *52.9* | 8 | 980 | 87.7 | 3 | 130 | 160 | 1 | 80 | 142.4 |
| Total MPs | 1185 | | | 262 | | | 726 | | | 522 | | | 50 | | | 33 | |
| Median selected slides | 1355 | | | 205 | | | 775 | | | 470 | | | 50 | | | 40 | |
| Recovery median (%) | 114.3 | | | 78.2 | | | 106.7 | | | 90 | | | 100 | | | 121.2 | |
| Average selected slides | 1240 | | | 225 | | | 726 | | | 488 | | | 52 | | | 40 | |
| RSD (%) | 41 | | | 44.2 | | | 39.3 | | | 7.2 | | | 42.2 | | | 40.8 | |
| Recovery average (%) | 104.6 | | | 85.9 | | | 104.7 | | | 93.4 | | | 105 | | | 121.2 | |
| Error (%) | 4,6% | | | -14,1% | | | 4,7% | | | 6,6% | | | 5% | | | 21,2% | |

In the next subsections (3.2., 3.3. and 3.4) the recoveries will be evaluated as a function of the ISO-defined size ranges for particles and the identified polymer type. To simplify the explanations colour codes are shown in the figures: Red indicates that the sample-based subsampling strategy does not detect a given size or polymer; orange denotes that the reference value (which was the value derived from the analysis of the overall slide) is not contained in the 99.7 % confidence interval associated to the experimental extrapolation (considering the 4 selected slices, $\overline{X} \pm 3 \cdot SD$ ); yellow, green and blue indicate statistical coincidence at 99.7 %, 95.5 % ($\overline{X} \pm 2 \cdot SD$) and 68.3 % ($\overline{X} \pm SD$) confidence levels, respectively (these confidence levels are usually referred to as the 99 %, 95 % and 68 % ones, respectively).

## 3.2. Evaluation of the predictions of MPs as a function of size

Our results will discriminate fibres and fragments, although this issue had not been reported previously (to the best of our knowledge) because most authors considered either total particles or just fragments alone. The only exception was the study from Thaysen et al. [17] who found that some of the strategies led to >50 % errors (measured as coefficient of variation, or RSD) when studying fibres, although one based on random squares along concentric circles of a filter achieved ca. 90 % recovery for pure water spiked with ca. 376 fibres.

**Figure 2** shows the recoveries obtained for samples 1 to 4 and the two blanks as a function of the size range. It is seen that most 68 % confidence intervals derived from the subsampling approach include the reference values, for both the total number of fibres and fragments. These results are encouraging because the recoveries for total fibres and fragments are close to 100 %, but for fibres in a sample and a blank (they two had a very sparse distribution of particles, which may yield problems as explained above). Furthermore, the 50-20 µm range presents excellent recoveries, which is important because that range contains usually most of the particles in the environmental samples. Detailed results from the entire characterization of the slides for each sample/blank can be seen in **Figures SM1, SM3, SM5, SM7, SM9 and SM11,** Supplementary Material. Detailed results for the average and 68 % confidence intervals gathered from the subsampling of the four selected slices of each sample/blank can be seen in **Figures SM2, SM4, SM6, SM8, SM10 and SM12,** Supplementary Material.

A curious result is that, although the estimated number of fragments in the 100-50 µm range for Blank #1 presents only a 25 % error (recovery = 125 %), the reference value is out of the confidence limits (**Figure 2**). This happens because SD equals to 0 (See **Figure SM10**, sum of fragments, 100-50 µm, supplementary material). Blank #2 has only 33 MPs and so it is difficult to get good results. Predictions for fibres were not satisfactory, as the full sample contained only 1 fibre (**Figure SM11**, supplementary material) while on the contrary, and positively enough, recoveries for fragments are good, with a maximum 25 % relative error.

| | % RECOVERIES | | | | | | | | | | | |
| | FIBRE | | | | | | FRAGMENTS | | | | | |
| | 5000-1000 | 1000-500 | 500-100 | 100-50 | 50-20 | TOTAL FIBRE | 5000-1000 | 1000-500 | 500-100 | 100-50 | 50-20 | TOTAL PARTICLES |
| Sample 1 | 250.0 | 50.0 | 125.0 | 0.0 | | 111.8 | | 250.0 | 102.3 | 108.4 | 103.3 | 104.4 |
| Sample 2 | 83.3 | 62.5 | 17.9 | | | 40 | 0.0 | 0.0 | 47.6 | 84.3 | 105.2 | 91.0 |
| Sample 3 | 125.0 | 90.9 | 133.3 | 250.0 | | 125.0 | | | 12.5 | 97.9 | 107.9 | 103.3 |
| Sample 4 | 250.0 | 200.0 | 76.9 | 125.0 | | 107.1 | 250.0 | 83.3 | 39.1 | 78.1 | 100.7 | 93.4 |
| Blank 1 | 250 | | 83 | 0 | | 100 | | | 167 | 125 | 89 | 105.0 |
| Blank 2 | | | 0 | | | 0.0 | | | 0 | 42 | 150 | 125 |

**Figure 2:** Recoveries (given as %) obtained for the sample-based subsampling strategy when the number of MPs is considered as a function of the size range. See section 3.1 for colour explanations.

## 3.3. Evaluation of individual polymers

It was seen above (section 3.2.) that it is possible to select four slices that represent very well the overall number of particles and the total number of MPs in the samples. A step forward is to study how well the different polymers are estimated from the predicted quantities of MPs (by subsampling). This was not considered too frequently in published subsampling studies as they focused on evaluating the total numbers of particles and/or MPs. The only exception we found is a recent paper [20] that considered the individual polymers to decide on the filter area leading to the best trade-off between workload and overall error. Despite the proposed solution was not the best in terms of accuracy for all polymers the authors restricted the measurement region to avoid unpractical working times for monitoring purposes.

It is true that differentiating polymer types increases the complexity of decision-making. To avoid major errors it was suggested to increase the overall number of particles and/or subareas when polymers are to be distinguished [19]. However, an advantage of our approach is that we do not restrict our work to 12.5 % of the filter as in that study. Hence, can 40 % of the area be enough to get acceptable results by LDIR? The results shown below point towards a positive answer.

**Figure 3** details the recoveries obtained per sample and polymer. Results are very good in general, as the 68 % confidence intervals of most predictions include the reference value (blue colour in the figures). Best recoveries occur in general for PP because it is the most abundant polymer and, so, subsampling captures best the overall distribution. It is worth noting that recoveries denoted as "0" correspond to polymers with very few items in the holder, typically 1 or 2 MPs (this can be verified in the corresponding tables at the supplementary material). Despite some recoveries are far from 100 % the confidence intervals include the reference values because of the relatively large standard deviation derived from the four estimations. Fibres have slightly more misidentifications but nevertheless in most cases the values derived from subsampling include the reference values within

76 the 68 % confidence intervals. Since in most occasions 95 % and 99 % confidence intervals are
77 used to report analytical values, this is a positive result.

| Polymer | FIBRES SAMPLE 1 | SAMPLE 2 | SAMPLE 3 | SAMPLE 4 | BLANK 1 | BLANK 2 | FRAGMENTS SAMPLE 1 | SAMPLE 2 | SAMPLE 3 | SAMPLE 4 | BLANK 1 | BLANK 2 |
|---|---|---|---|---|---|---|---|---|---|---|---|---|
| Acrylate | 250.0 | | | 0.0 | | | 0.0 | | | 83.3 | | |
| Alkyd Varnish | | | | | | | 142.9 | 0.0 | 83.3 | | | |
| EVA | | | | | | | 0.0 | | | | | |
| PA | | 250.0 | | | | | 25.0 | 93.8 | 100.7 | 50.0 | | 156.3 |
| PE | | 500.0 | 250.0 | 0.0 | | | 100.0 | 50.0 | 47.6 | 117.2 | | |
| PC | | | | | | | | | | | | |
| PET | 102.3 | 0.0 | 150.0 | 120.0 | | 0.0 | 96.8 | 76.9 | 107.6 | 75.9 | 250.0 | 50.0 |
| PMMA | | | | | | | | 250.0 | | | | |
| POM | | | | | | | | | | 0.0 | | 0.0 |
| PP | 50.0 | 0.0 | 83.3 | 93.8 | 100.0 | | 114.5 | 102.7 | 104.2 | 93.1 | 125.0 | 125.0 |
| PS | 150.0 | 83.3 | 250.0 | | | | 105.9 | 95.5 | 127.8 | 134.6 | 0.0 | |
| PU | | | | | | | 70.3 | 83.3 | 83.3 | 0.0 | 250.0 | |
| PVC | | 0.0 | | | | | 187.5 | 50.0 | 0.0 | 83.3 | | |
| Tyre | | | | | | | 250.0 | | | | 62.5 | |
| ABS | | | 0.0 | | | | 100.0 | 83.3 | 108.5 | 159.1 | | 214.3 |
| rubber | | | | | | | 47.1 | 166.7 | 83.3 | 125.0 | 104.2 | 83.3 |
| PTFE | | | | | | | 250.0 | | | 0.0 | | |
| TOTAL | 111.8 | 40.0 | 125.0 | 107.1 | 100.0 | 0.0 | 104.4 | 91.1 | 103.3 | 92.4 | 105.6 | 125.0 |

78

79 **Figure 3:** Recoveries (given as %) obtained using the sample-based subsampling strategy when
80 the number of MPs is considered as function of the polymer, see section 3.1 for colour
81 explanations.

82

## 3.4. Evaluation of both polymer and size

84 Next, it is worth studying how the predictions are when both polymer type and size are considered.
85 It is expected that the results will be much more complex than for the two previous studies because
86 we are not extrapolating to a grand total (e.g., the overall number of MPs) but to subtotals where the
87 "true" number of particles is quite low and, accordingly, they are distributed (very) unevenly in the
88 sample holder. Results obtained for the samples and the blanks are given in **Figure 4** where the
89 predictions are detailed by polymer, size and type of MP (fibre vs fragment).

90 Two different situations can be observed. As somehow expected, fibres yield less satisfactory results
91 because their abundance is very low (hence, their sparsity is high and their distribution not easy to
92 capture with a reduced number of subareas). So, even when the sample-based approach accurately
93 predicted the overall number of items per size (**Figure 2**) and the polymers (**Figure 3**) further
94 subdivisions yield unsatisfactory results. In case the samples contain hundreds of fibres (as when
95 clothes washing is studied [25]) subsampling will perform better as it will recover best the underlying
96 distribution of particles. But we had not such kind of samples.

97  Up to 16 cells (i.e., combinations of size and polymer type for fibres) out of 39 in **Figure 4** reported
98  no counts for fibres, despite they existed in the samples (although only with as low quantities as 1
99  or 2 items; see **Figures SM1, SM3, SM5, SM7, SM9 and SM11** in Supplementary Material). The
100 other predictions included the reference value in the confidence intervals.

101 When fragments are considered, larger particles (5000-500 µm) yielded worse results, again due to
102 their sparse presence throughout the reflective slide. An alternative would be to characterize big
103 particles (>500 µm) using traditional FTIR-ATR. If they are to be studied by LDIR we suggest setting
104 the system to detect and analyse only that range and scan all the plate. This is really very fast (for
105 their amount be usually very low) and no subsampling is required, so full representativeness will be
106 obtained, and errors avoided.

107 In the 100-50 µm size range the sample-based subsampling led to good predictions as 32 out of 40
108 cases (80 %) included the reference values within the 68 % experimental confidence intervals. Only
109 8 cases (20 %) yielded no predictions, of which 4 corresponded to the same sample. Results were
110 much better for <100 µm particles as their abundance is higher in the samples, and predictions are
111 very good for the smallest range (50-20 µm), with most predictions including the reference value in
112 their 68 % confidence intervals. With regard to polymer types, we have not observed systematic
113 errors for a particular type, further from their larger or lower presence in the samples, which is an
114 uncontrollable factor because all of them were field samples. Only when the reference value was 1
115 or 2 items for a specific size and/or polymer no predictions were obtained in some cases (11 out of
116 59 situations, 19 % of the overall predictions in **Figure 4**).

| X ± SD (68% SC.) | X ± 2SD (95% S.C) | X ± 3SD (99% S.C) | OUT OF 99% S.C | NOT PRESENT |
|---|---|---|---|---|

**SAMPLE 1**

| | FIBRES | | | | | FRAGMENTS | | | | |
|---|---|---|---|---|---|---|---|---|---|---|
| | 5000-1000 | 1000-500 | 500-100 | 100-50 | 50-20 | 5000-1000 | 1000-500 | 500-100 | 100-50 | 50-20 |
| Acrylate | | | 250.0 | | | | | | | 0.0 |
| Alkyd Varnish | | | | | | | | | | 142.9 |
| EVA | | | | | | | | | | 0.0 |
| PA | | | | | | | | | | 25.0 |
| PE | | | | | | | | | 0.0 | 111.1 |
| PC | | | | | | | | | | |
| PET | | 0.0 | 132.4 | 0.0 | | | | 41.7 | 150.0 | 100.0 |
| PMMA | | | | | | | | | | |
| POM | | | | | | | | | | |
| PP | | | 50.0 | | | | 250.0 | 166.7 | 113.6 | 112.6 |
| PS | 250.0 | 250.0 | 142.9 | 0.0 | | | 250.0 | 101.9 | 102.0 | 106.3 |
| PU | | | | | | | | 0.0 | 83.3 | 74.1 |
| PVC | | | | | | | | | 250.0 | 166.7 |
| Tyre | | | | | | | | 250.0 | 250.0 | 250.0 |
| ABS | | | | | | | | | | 100.0 |
| rubber | | | | | | | | 0.0 | 44.1 | 48.5 |
| PTFE | | | | | | | | | 250.0 | |

**SAMPLE 2**

| | FIBRES | | | | | FRAGMENTS | | | | |
|---|---|---|---|---|---|---|---|---|---|---|
| | 5000-1000 | 1000-500 | 500-100 | 100-50 | 50-20 | 5000-1000 | 1000-500 | 500-100 | 100-50 | 50-20 |
| Acrylate | | | | | | | | | | |
| Alkyd Varnish | | | | | | | | | | 0.0 |
| EVA | | | | | | | | | | |
| PA | | | 250.0 | | | | | | 125.0 | 90.9 |
| PE | | | 0.0 | | | | | 62.5 | 62.5 | 35.7 |
| PC | | | | | | | | | | |
| PET | | 0.0 | 0.0 | | | | | 0.0 | 83.3 | 75.0 |
| PMMA | | | | | | | | | 250.0 | 250.0 |
| POM | | | | | | | | | | |
| PP | | 0.0 | 0.0 | | | | | 125.0 | 58.8 | 121.4 |
| PS | 125.0 | 0.0 | | | | 0.0 | 0.0 | 35.7 | 85.4 | 125.0 |
| PU | | | | | | | | | 0.0 | 125.0 |
| PVC | 0.0 | | | | | | | 0.0 | 125.0 | 83.3 |
| Tyre | | | | | | | | | | |
| ABS | | | | | | | | | 125.0 | 62.5 |
| rubber | | | | | | | | | 250.0 | 125.0 |
| PTFE | | | | | | | | | | |

**SAMPLE 3**

| | FIBRES | | | | | FRAGMENTS | | | | |
|---|---|---|---|---|---|---|---|---|---|---|
| | 5000-1000 | 1000-500 | 500-100 | 100-50 | 50-20 | 5000-1000 | 1000-500 | 500-100 | 100-50 | 50-20 |
| Acrylate | | | | | | | | | | |
| Alkyd Varnish | | | | | | | | | | 83.3 |
| EVA | | | | | | | | | | |
| PA | | | | | | | | | 83.3 | 101.1 |
| PE | | | 250.0 | | | | | 0.0 | 35.7 | 57.7 |
| PC | | | | | | | | | | |
| PET | | 125.0 | 156.3 | 250.0 | | | | 0.0 | 83.3 | 122.8 |
| PMMA | | | | | | | | | | |
| POM | | | | | | | | | | 0.0 |
| PP | 125.0 | 0.0 | 90.9 | | | | | 19.2 | 100.6 | 110.3 |
| PS | | | 250.0 | | | | | 0.0 | 125.0 | 131.3 |
| PU | | | | | | | | | | 83.3 |
| PVC | | | | | | | | | | 0.0 |
| Tyre | | | | | | | | | | |
| ABS | | | 0.0 | | | | | 0.0 | 166.7 | 103.3 |
| rubber | | | | | | | | | | 83.3 |
| PTFE | | | | | | | | | | |

117

**Figure 4**. Recoveries (given as %) as a function of polymer and size, see section 3.1 for colour explanations.

Legend:
- x ± SD (68% SC.)
- x ± 2SD (95% S.C)
- x ± 3SD (99% S.C)
- OUT OF 99% S.C.
- NOT PRESENT

| Polymer | FIBRES | | | | | FRAGMENTS | | | | |
|---|---|---|---|---|---|---|---|---|---|---|
| | 5000-1000 | 1000-500 | 500-100 | 100-50 | 50-20 | 5000-1000 | 1000-500 | 500-100 | 100-50 | 50-20 |
| **SAMPLE 4** | | | | | | | | | | |
| Acrylate | | | 0.0 | | | | | 250.0 | 0.0 | 0.0 |
| Alkyd | | | | | | | | | | |
| Varnish | | | | | | | | | | |
| EVA | | | | | | | | | | |
| PA | | | | | | | | | 83.3 | 34.1 |
| PE | | | 0.0 | | | | | 0.0 | 62.5 | 142.9 |
| PC | | | | | | | | | | |
| PET | 250.0 | 200.0 | 92.1 | | | | | 16.7 | 79.5 | 89.3 |
| PMMA | | | | | | | | | | |
| POM | | | | | | | | | | |
| PP | 250.0 | | 50.0 | 125.0 | | 250.0 | 83.3 | 57.7 | 83.3 | 93.6 |
| PS | | | | | | | | | 0.0 | 159.1 |
| PU | | | | | | | | | | 0.0 |
| PVC | | | | | | | | 0.0 | | 125.0 |
| Tyre | | | | | | | | | | |
| ABS | | | | | | | | | 250.0 | 214.3 |
| rubber | | | | | | | | | | 104.2 |
| PTFE | | | | | | | | 0.0 | | |
| **BLANK 1** | | | | | | | | | | |
| Acrylate | | | | | | | | | | |
| Alkyd | | | | | | | | | | |
| Varnish | | | | | | | | | | |
| EVA | | | | | | | | | | |
| PA | | | | | | | | | | |
| PE | | | | | | | | | | |
| PC | | | | | | | | | | |
| PET | | | | | | | | 250.0 | | |
| PMMA | | | | | | | | | | |
| POM | | | | | | | | | | |
| PP | 250.0 | | 83.3 | 0.0 | | | | 250.0 | 250.0 | 100.0 |
| PS | | | | | | | | | | 0.0 |
| PU | | | | | | | | | | 250.0 |
| PVC | | | | | | | | | | |
| Tyre | | | | | | | | 0.0 | 125.0 | 55.6 |
| ABS | | | | | | | | | | |
| rubber | | | | | | | | 166.7 | 62.5 | 100.0 |
| PTFE | | | | | | | | | | |
| **BLANK 2** | | | | | | | | | | |
| Acrylate | | | | | | | | | | |
| Alkyd | | | | | | | | | | |
| Varnish | | | | | | | | | | |
| EVA | | | | | | | | | | |
| PA | | | | | | | | | | 156.3 |
| PE | | | | | | | | | | |
| PC | | | | | | | | | | |
| PET | | | 0.0 | | | | | 0.0 | 0.0 | 250.0 |
| PMMA | | | | | | | | | | |
| POM | | | | | | | | | | 0.0 |
| PP | | | | | | | | | 250 | 0.0 |
| PS | | | | | | | | | | |
| PU | | | | | | | | | | |
| PVC | | | | | | | | | | |
| Tyre | | | | | | | | | | |
| ABS | | | | | | | | | | 214.3 |
| rubber | | | | | | | | | 0 | 107.1 |
| PTFE | | | | | | | | | | |

120

121

122 **Figure 4** (continues):

123

124

## 3.5. Comparison with other subsampling approaches for reflective slides

126 For the sake of comparison other *ad-hoc* strategies to sample reflective slides were studied. They
127 can be considered as "deterministic" or systematic in the sense that a predefined, fixed distribution
128 of subareas is scheduled in advance and applied to the samples, without a preliminary counting. We
129 fixed the characterization area to be 50 % of the reflective slide following previous recommendations
130 [4]. This allowed us to set very simple and practical divisions of the plate, avoiding the need to carry
131 out further selection on the slides (which would yield the sample-based study itself!). All the particles
132 in these regions were characterized and evaluated for MPs. We acknowledge that this additional 10
133 % area may complicate the straightforward comparison with the previous approach, although we feel
134 that this is not a relevant problem here, as measuring 50 % of the plate can be considered as a
135 benchmark, being considered as a statistically reliable option [4].

136 Several options were considered (refer to **Figure 1** for their visualization): (i) measuring half the
137 reflective slide "horizontally" (slices (columns) 1-10 and rows A-E), (ii) "horizontal alternating",
138 consisting of measuring half of the reflective slide but alternating rows (odd and even –all slices-),
139 (iii) "vertical"; i.e., measuring half of the reflective slide "vertically" (slices 1-5 –all rows-), (iv) "vertical
140 alternating": involves measuring half of the reflective slide but alternating slices (slices 1, 3, 5, 7 and
141 9 –all rows-). The results gathered from these alternatives will be given only for sample #1 (33,032
142 total detected particles, measured particle sizes >20 µm and 1,185 identified MPs) as this was
143 considered an exemplary sample.

144 It can be seen that, in general, the systematic approaches do not outperform the sample-based
145 strategy to evaluate the number of particles, even when up to 50 % of the reflective slide was
146 considered, see **Table 3**. The two complementary halves of the plate (i.e., odd and even slices –or
147 horizontal- rows) show logically complementary errors for the evaluation of the total number of
148 particles and MPs.

## 3.6. Application to gold-coated filters

150 The sample-based approach was applied also to the other relevant mode for deploying the extracted
151 particles on a reflective plate: gold-coated filters. The overall working process is faster than for Kevley
152 reflective slides because the total surface of a 25 mm diameter filter is about one third that of a
153 reflective slide. The results obtained for samples #5 and #6 are resumed in **Tables 4 and 5,** and
154 **Figure 5**. **Table 4** reveals that the prediction of the total number of particles is very good, with
155 excellent recoveries and small relative standard deviations (RSDs < 10 %), and very slight
156 overestimations (< 10 %). The prediction of the total number of MPs (**Table 5**) is also good, within
157 low variability (RSD < 20 %) and very good recoveries, close to 100 % (ca. 107 % and 112 %).

158

| Systematic subsample approach | Relative errors (%) | |
|---|---|---|
| | Total particles | Total MPs |
| Horizontal (all slices, rows A-E) | 15 | 13 |
| Horizontal (all slices, rows F-J) | -15 | -13 |
| Horizontal alternating (rows A, C, E, G and I, all slices) | 10 | 5 |
| Horizontal alternating (rows B, D, F, H and J, all slices) | -10 | -5 |
| Vertical (slices 1-5, all rows) | 15 | 9 |
| Vertical (slices 6-10, all rows) | -15 | -9 |
| Vertical alternating (slices 1, 3, 5, 7 and 9, all rows) | 2 | 5 |
| Vertical alternating (slices 2, 4, 6, 8 and 10, all rows) | -2 | -5 |

159

160 **Table 3**: Relative errors (given as %) obtained when considering eight *ad-hoc* systematic strategies
161 to subsample reflective slides. In all cases 50 % of the total surface of the reflective slide was
162 analyzed for all particles. Refer to **Figure 1** to visualize the distribution of the subsampled areas.

163

164 Noticeably, in our studies with the gold-coated filters we experienced only recoveries higher than
165 100 %, not lower. A possible explanation for the slight overestimations is the existence of three
166 practical limitations:

167     i)     The placement of the filter in the dedicated holder disturbs its edge making it irregular
168         and, so, the border of the filter cannot be measured because the software will not be able
169         to properly focus the laser beam and errors will stop the system.

170     ii)    When defining the subareas, rectangles let small filter triangles unseleced below and
171         above them (see **Figure 1**) because the software does not allow to deploy trapezoids.
172         Also, take into account the previous limitation when setting the subareas.

173     iii)   Opposed to what happens for the Kevley slides, each slice does not represent 10 % of
174         the surface (see **Figure 1**) and, so, geometric factors need to be used for extrapolation.
175         They are: slices 1 & 10 = 4.1 % of the filter; slices 2 & 9 = 8.1 %; slices 3 & 8 = 10 %;
176         slices 4 & 7 = 11.1 %; slices 5 & 6 = 11.8 %. In total, they comprise 90.2 % of the filter.

177 As for the studies above, recoveries when the total number of MPs were classified as a function of
178 size (**Figure 5**) and polymer types (**Figure 6**) resulted very satisfactory. The recoveries for the total
179 amount of fibres and fragments are very close to 100 % (**Table 4**, columns showing total amounts;
180 **Table 5**, last row). However, when only one or two particles of a given size are present on the whole
181 filter, predictions are not good (in these examples, for fibres) because it is difficult to capture
182 statistically their very sparse distribution.

183 Detailed results from the entire characterization of the slides for each sample can be seen in **Figures**
184 **SM13 and SM15,** whereas detailed results for the average and 68 % confidence intervals gathered from the subsampling of the four selected slices of each sample are shown in **Figures SM14 and SM16,** Supplementary Material.

**Table 4**: Evaluation of the total number of particles in a gold-coated filter by means of sample-based subsampling. The cells corresponding to the selected slices have a grey background. RSD stands for relative standard deviation of the mean, given as a percentage.

| | Sample 5 | | | Sample 6 | |
|---|---|---|---|---|---|
| Slice # | Total particles | % error | Slice # | Total particles | % error |
| 2 | 2605 | -17.2 | 10 | 1659 | -48.7 |
| 9 | 2667 | -15.3 | 9 | 3148 | -2.5 |
| 1 | 2805 | -10.9 | 7 | 3252 | 0.7 |
| *5* | *2932* | *-6.8* | *6* | *3373* | *4.4* |
| *10* | *3073* | *-2.3* | *5* | *3424* | *6.0* |
| *6* | *3458* | *9.9* | *8* | *3430* | *6.2* |
| *3* | *3560* | *13.1* | *1* | *3610* | *11.8* |
| 7 | 3838 | 22.0 | 4 | 3694 | 14.4 |
| 4 | 3856 | 22.5 | 3 | 4100 | 26.9 |
| 8 | 4110 | 30.6 | 2 | 4111 | 27.3 |
| Total particles (100 % filter) | 3147 | | 3230 | | |
| Median selected slides | 3265 | | 3427 | | |
| Recovery median (%) | 103.8 | | 106.1 | | |
| Average selected slides | 3256 | | 3459 | | |
| RSD (%) | 9.2 | | 3.0 | | |
| Recovery average | 103.5 | | 107.1 | | |

207

208 **Table 5**: Evaluation of the total number of microplastics in a gold-coated filter by means of sample-
209 based subsampling. The cells corresponding to the selected slices (**Table 4**) have a grey
210 background. RSD stands for relative standard deviation of the mean, given as a percentage.

| Sample 5 | | | Sample 6 | | |
|---|---|---|---|---|---|
| Slice # | Total Microplastics | % error | Slice # | Total microplastics | % error |
| 2 | 195 | -27.7 | 10 | 268 | -4.4 |
| 1 | 198 | -26.8 | 9 | 333 | -2.3 |
| *5* | *246* | *-9.0* | *6* | *390* | *-0.4* |
| *6* | *280* | *3.6* | *5* | *415* | *0.4* |
| 7 | 288 | 6.8 | 2 | 420 | 0.6 |
| 4 | 288 | 6.8 | *8* | *420* | *0.6* |
| 8 | 300 | 11.1 | 7 | 477 | 2.4 |
| *3* | *310* | *14.8* | 4 | 495 | 3.0 |
| *10* | *317* | *17.4* | 3 | 530 | 4.2 |
| 9 | 383 | 41.7 | *1* | *585* | *6.0* |
| Total MPs (100 % filter) | 270 | | 403 | | |
| Median selected slides | 295 | | 418 | | |
| Recovery median (%) | 109.2 | | 103.6 | | |
| Average selected slides | 288 | | 453 | | |
| RSD (%) | 11.2 | | 19.7 | | |
| Recovery average | 106.7 | | 112.3 | | |

211

212 For instance, Sample #5 has only 1 fibre in the 5000-1000 µm range (**Figure SM13, Supplementary**
213 **Material**) which was not captured when subsampling, and Sample#6 has 3 fibres –also undetected-
214 in the 1000-500 µm range (**Figure SM15, Supplementary Material**). When the classification of the
215 MPs is done considering the polymers (**Figure 6**) there was good agreement between the
216 extrapolated and the reference values for the most abundant polymers (Sample #5: PE, PA, PS and
217 PP; Sample #6: PE and PP). Those predictions out of the 99 % confidence intervals in **Figure 6** are
218 caused, again, by the presence of very few particles (only one) of those sizes in the overall filter (see
219 **Figures SM13 and SM15**, **Supplementary Material**).

220

| % RECOVERIES | | | | | | | | | | | |
| FIBRE | | | | | | FRAGMENT | | | | | |
| Size | 5000-1000 | 1000-500 | 500-100 | 100-50 | 50-20 | TOTAL FIBRE | 5000-1000 | 1000-500 | 500-100 | 100-50 | 50-20 | TOTAL FRAGMENT |
| SAMPLE 5 | 0.0 | 57.8 | 108.8 | 85.0 | | 88.3 | | | 235.0 | 107.7 | 104.9 | 109.3 |
| SAMPLE 6 | 212.5 | 0.0 | 114.5 | 112.5 | | 107.9 | | | 98.4 | 103.1 | 118.1 | 113.0 |

**Figure 5**: Evaluation of the total amount of MPs considering size (the values represent the recoveries, given as %). Blue =statistical coincidence with reference value (68 % confidence level), red = no identification when subsampling.

| | % RECOVERY | | | |
| Polymer | FIBRES | | FRAGMENTS | |
| | SAMPLE 5 | SAMPLE 6 | SAMPLE 5 | SAMPLE 6 |
| Acrylate | | 212.5 | 0.0 | 0.0 |
| Alkyd Varnish | | | | |
| EVA | 250.0 | | | 0.0 |
| PA | | 212.5 | 92.5 | 205.6 |
| PE | 35.4 | 41.9 | 104.9 | 111.1 |
| PC | | | 0.0 | |
| PET | 106.3 | 245.6 | 189.5 | 85.0 |
| PMMA | | | 0.0 | 106.3 |
| Polyacetal or POM | | | | |
| PP | 102.8 | 103.8 | 109.6 | 114.2 |
| PS | 106.3 | | 155.9 | 212.5 |
| PU | 0.0 | | 115.6 | 219.3 |
| PVC | | | 0.0 | 0.0 |
| Tyre | | | | |
| ABS | | | 77.1 | 70.8 |
| rubber | | | | |
| PTFE | | | | |
| TOTAL | 88.3 | 107.9 | 109.3 | 113.0 |

**Figure 6**: Evaluation of the total amount of MPs considering type of polymer (the values represent the recoveries, given as %), see section 3.1 for colour explanations.

As for the Kevley slides, more problems appear when both the size and polymer of the MPs are estimated disaggregated (**Figure SM17**), mostly for Sample #5 because it has a lower number of MPs than Sample #6 (270 vs 403, see **Figures SM13 and SM15**, **Supplementary Material**) which yields less MPs of a particular size and of a particular polymer. Despite this, the overall estimations seem quite acceptable, although in the particular example some overestimations are seen for fragments of Sample #6 (and one case for Sample #5), with too high overestimations (recoveries reaching 600 %). This is due to the fact that only one particle of these types was present in the overall filter, and since it was recorded in the subsampling, the final extrapolation led to an overestimation. Note that the situations that led to extreme errors were caused by the same problem:

238 a very reduced number of particles of a given class. It they are unnoticed in the subsampling stage,
239 underestimations may occur, otherwise, overestimations appear.

240

241 **3.7. Operational considerations**

242 It is worth noting that sample-based subsampling was found to show its maximum efficiency (i.e.,
243 savings on time and staff dedication) when the samples contain many particles, typically >15,000 -
244 18,000. Recall that, as explained in the previous sections, blanks do not need to be subsampled
245 because they contain so few particles that the savings on time are not really worth it and, besides,
246 the sparsity of the particles lead to unnecessarily high errors.

247 In our experience, high loads of particles in the gold-coated filters should be avoided not only to
248 avoid filter clogging but superimposed particles. Likely this means that less particles will be present
249 in a gold-coated filter than in a Kevley slide. If the sample contains too many particles it is preferable
250 to use more filters or to reduce the measuring aliquot. None of them is a good solution because of
251 the cost of the filters (and the time needed to measure them) and the risks of extrapolating from
252 small aliquots to the overall sample. This is a tough trade-off that must be undergone in each
253 laboratory for each study, likely with no optimal solution.

254 Staff dedication can be summarized as follows 1$^{st}$ step: to set slices and order automatic counting of
255 total number of particles on them (dedication: 30 min per slice, total ca. 3 h although dedication is
256 required only after the system counted the particles of each slice); 2$^{nd}$ step: selection of the slices
257 (15 min); 3$^{rd}$ step: characterization of all particles in the 4 selected slices, automatic without staff
258 dedication.

259 Thus, for a gold-coated filter containing 8,000-9,000 particles (those that usually can measure a
260 LDIR/24 h) the sample-based strategy will need 2 h of staff dedication to mark the slices, make the
261 calculations and select 4 of them, plus another ca. 10 h for their automatic measurement. A possibility
262 to reduce the staff workload a little more is to discard slices 1 and 10 from scratch because they
263 usually lead to clear underestimations and, so, consider only the 8 remaining ones. That would be a
264 pragmatic solution but statistically limits the possibility that any particle would be measured and we
265 have seen that sometimes these slices are important (see, e.g. **Table 4**).

266

## 4. CONCLUSIONS

A sample-based subsampling strategy especially suited for the new and fast quantum-cascade laser-based reflectance-transflectance imaging system from Agilent (commercially known as 8700 LDIR) has been presented in this work. It allows the analyst to decide the overall size of the scanning area that may be adequate for its particular problem by studying the error when evaluating the number of particles, as this acts as a proxy for the error when assessing the total number of microplastics. The approach is applied *ex-ante* the infrared characterization of the particles and, therefore, constitutes a way to reduce the workload to analyze environmental samples when large numbers of particles are present on Kevley slides or reflective (gold-coated) filters. We found that considering 40 % of the overall reflective surface results in a good trade-off between workload and accuracy. The time required for the analysis of a Kevley slide with ca. 20,000 particles and for a gold-coated filter with ca. 9,000 particles can be reduced by 40-50 %.

The approach has proved effective across several key environmental matrices. It predicts reliably the total number of particles (errors as average recoveries < 5-10%) and microplastics (errors <20 %) in the 5000-20 µm range. It performs also very well when estimating the number of MPs either by polymer or size range (in many cases with almost 100 % average recoveries). However, when only one or two particles of a given size and/or polymer are present in the overall holder, over- or under-estimations can occur depending on whether they were detected in the subsampling process because the final values proceed from extrapolation of the measured subareas. Typically this occurs for relatively big particles, > 500 µm, as they are not too frequent in the samples and their distribution is very sparse. When the number of MPs are disaggregated by both polymer type and particle size, some challenges arise. In particular with very sparse distributions (usually large particles) due to under- or over-estimation, as commented above.

This explains why it is not recommended to subsample the blanks, as it does not save much time and the errors on the estimations are large, because of the sparseness of the particles in all size ranges.

Despite those problems, common to all subsampling strategies, the approach performs notably well for particles in the 100-50 and 50-20 µm size ranges, especially for the most abundant polymers in the samples. It is worth remembering that over 60 % of MPs of a sample fall usually within the 50-20 µm range, and more than 80 % within the 100-20 µm one. In this fraction, the subsampling approach is highly accurate for the most abundant polymers (average recoveries close to 100 % and low RSD).

For studies that aim to provide detailed results in size ranges > 500 µm, two possibilities can be seen: i) analyzing the entire surface of the holder restricting the LDIR setup to detecting high sizes, as they constitute only a small fraction of the total MPs, being this a straightforward, time-efficient option, and ii) pick up those big particles manually with micro tweezers and characterize them by traditional FTIR-ATR, which is really very fast, so full representativeness will be obtained, and errors avoided.

To ensure successful disaggregated predictions (considering both size range and polymer) sample-based subsampling will benefit from evaluating samples containing high numbers of particles; for instance (and based on our empirical experience), around 19,000 particles on a reflective slide or 4,000 particles on a gold-coated filter. That would make the underlying distribution of MPs less sparse and, so, more easily detectable by subsampling.

## ACKNOWLEDGEMENTS

This work constitutes a part of the LABPLAS project (Grant Agreement No. 101003954), supported by the EU H2020 program. Funding to SplashMare project (PID2022-138421OB-C21, MICIU/AEI/10.13039/501100011033 and by FEDER, UE) is acknowledged. The Program "Consolidación e Estructuración de Unidades de Investigación Competitivas" of the Galician Government (Xunta de Galicia) is acknowledged (Grant ED431C 2021/56).

## CRediT authorship contribution statement

**Adrián López-Rosales**: Conceptualization, Methodology, Validation, Formal analysis, Investigation, Data Curation, Writing - Original Draft. **José Andrade**: Conceptualization, Writing - Review & Editing, Supervision. **Borja Ferreiro**: Investigation, Writing - Review & Editing. **Soledad Muniategui-Lorenzo**: Review & Editing, project administration, funding.

# SAMPLE-BASED SUBSAMPLING STRATEGIES TO IDENTIFY MICROPLASTICS IN THE PRESENCE OF A HIGH NUMBER OF PARTICLES USING QUANTUM-CASCADE LASER-BASED INFRARED IMAGING

Adrián López-Rosales, José M. Andrade*, Borja Ferreiro, Soledad Muniategui

Group of Applied Analytical Chemistry, Institute of Environmental Sciences (IUMA), University of A Coruña, Campus da Zapateira s/n, E-15071, A Coruña, Spain

(*) andrade@udc.es, fax: +34981167065

# SUPPLEMENTARY MATERIAL

The supplementary material includes 17 figures detailing the number of microplastics identified in the samples and blanks under study considering both a full-scanning of the plate or filter and the sample-based subsampling strategy (for which 4 slices were selected). For the latter, averages and standard deviations are shown.

See Main text for a description of the samples.

**NOTE:** The last row of even tables (SM2 … 16) show the average and standard deviation of the results derived from the four slices separately (they are not the mere sum of the upper results shown in the tables)

448

449

450    .Figure SM1. Sample 1 (Entire characterization)

| Polymer | FIBRE 5000-1000 | 1000-500 | 500-100 | 100-50 | 50-20 | FRAGMENT 5000-1000 | 1000-500 | 500-100 | 100-50 | 50-20 | Polymer | TOTAL | FIBRES | PARTICLES |
|---|---|---|---|---|---|---|---|---|---|---|---|---|---|---|
| Acrylate | | | 1 | | | | | | | 1 | Acrylate | 2,0 | 1 | 1 |
| Alkyd Varnish | | | | | | | | | | 7 | Alkyd Varnish | 7,0 | 0 | 7 |
| EVA | | | | | | | | | | 1 | EVA | 1,0 | 0 | 1 |
| PA | | | | | | | | | | 10 | PA | 10,0 | 0 | 10 |
| PE | | | | | | | | | 4 | 36 | PE | 40,0 | 0 | 40 |
| PC | | | | | | | | | | | PC | 0,0 | 0 | 0 |
| PET | | 4 | 17 | 1 | | | | 6 | 5 | 20 | PET | 53,0 | 22 | 31 |
| PMMA | | | | | | | | | | | PMMA | 0,0 | 0 | 0 |
| Polyacetal or POM | | | | | | | | | | | POM | 0,0 | 0 | 0 |
| PP | | | 5 | | | | 1 | 6 | 33 | 211 | PP | 256,0 | 5 | 251 |
| PS | 1 | 1 | 7 | 1 | | | 2 | 27 | 98 | 529 | PS | 666,0 | 10 | 656 |
| PU | | | | | | | | 2 | 3 | 27 | PU | 32,0 | 0 | 32 |
| PVC | | | | | | | | | 2 | 6 | PVC | 8,0 | 0 | 8 |
| Tyre | | | | | | | | 2 | 10 | 7 | Tyre | 19,0 | 0 | 19 |
| ABS | | | | | | | | | | 5 | ABS | 5,0 | 0 | 5 |
| rubber | | | | | | | | 1 | 17 | 67 | rubber | 85,0 | 0 | 85 |
| PTFE | | | | | | | | | 1 | | PTFE | 1,0 | 0 | 1 |
| | | | | | | | | | | | | 0,0 | 0 | 0 |
| | | | | | | | | | | | | 0,0 | 0 | 0 |
| Suma | 1,0 | 5,0 | 30,0 | 2,0 | 0,0 | 0,0 | 3,0 | 44,0 | 173,0 | 927,0 | | 1185,0 | 38 | 1147 |

451

452

453    **Figure SM2. Sample 1 (Subsampling of 4 slices, mean ± standard deviation)**

| Polymer | FIBRE 5000-1000 | 1000-500 | 500-100 | 100-50 | 50-20 | PARTICLE 5000-1000 | 1000-500 | 500-100 | 100-50 | 50-20 | Polymer | TOTAL |
|---|---|---|---|---|---|---|---|---|---|---|---|---|
| Acrylate | | | 2,5 ± 5 | | | | | | | | Acrylate | 2,5 ± 5 |
| Alkyd Varnish | | | | | | | | | | 10 ± 8,1 | Alkyd Varnish | 10 ± 8,1 |
| EVA | | | | | | | | | | | EVA | |
| PA | | | | | | | | | | 2,5 ± 5 | PA | 2,5 ± 5 |
| PE | | | | | | | | | | 40 ± 8,1 | PE | 40 ± 8,1 |
| PC | | | | | | | | | | | PC | |
| PET | | | 22,5 ± 18,9 | | | | | 2,5 ± 5 | 7,5 ± 5 | 20 ± 16,3 | PET | 52,5 ± 27,5 |
| PMMA | | | | | | | | | | | PMMA | |
| Polyacetal or POM | | | | | | | | | | | POM | |
| PP | | | 2,5 ± 5 | | | | 2,5 ± 5 | 10 ± 8,1 | 37,5 ± 33 | 237,5 ± 136,9 | PP | 290 ± 154,9 |
| PS | 2,5 ± 5 | 2,5 ± 5 | 10 ± 11,5 | | | | 5 ± 10 | 27,5 ± 15 | 100 ± 57,7 | 562,5 ± 256,4 | PS | 710 ± 303,3 |
| PU | | | | | | | | | 2,5 ± 5 | 20 ± 27 | PU | 22,5 ± 32 |
| PVC | | | | | | | | | 5 ± 10 | 10 ± 8,1 | PVC | 15 ± 17,3 |
| Tyre | | | | | | | | 5 ± 10 | 25 ± 33,1 | 17,5 ± 23,6 | Tyre | 47,5 ± 56,1 |
| ABS | | | | | | | | | | 5 ± 5,7 | ABS | 5 ± 5,7 |
| rubber | | | | | | | | | 7,5 ± 9,5 | 32,5 ± 27,5 | rubber | 40 ± 35,5 |
| PTFE | | | | | | | | | 2,5 ± 5 | | PTFE | 2,5 ± 5 |
| Suma | 2,5 ± 5 | 2,5 ± 5 | 37,5 ± 30,9 | | | | 7,5 ± 9,5 | 45 ± 23,8 | 187,5 ± 68,4 | 957,5 ± 433,3 | | |

| Total fibres | 42,5 ± 34 | | Total particles | 1197,5 ± 485,5 | | Total | 1240 ± 508,3 |
|---|---|---|---|---|---|---|---|

454

455

456

457

458

459 **Figure SM3. Sample 2 (Entire characterization)**

| Polymer | FIBRE | | | | | FRAGMENT | | | | | Polymer | TOTAL | FIBRES | PARTICLES |
|---|---|---|---|---|---|---|---|---|---|---|---|---|---|---|
| | 5000-1000 | 1000-500 | 500-100 | 100-50 | 50-20 | 5000-1000 | 1000-500 | 500-100 | 100-50 | 50-20 | | | | |
| Acrylate | | | | | | | | | | | Acrylate | 0,0 | 0 | 0 |
| Alkyd Varnish | | | | | | | | | | 2 | Alkyd Varnish | 2,0 | 0 | 2 |
| EVA | | | | | | | | | | | EVA | 0,0 | 0 | 0 |
| PA | | | 1 | | | | | | 2 | 22 | PA | 25,0 | 1 | 24 |
| PE | | | 1 | | | | | 4 | 4 | 7 | PE | 16,0 | 1 | 15 |
| PC | | | | | | | | | | | PC | 0,0 | 0 | 0 |
| PET | | 6 | 11 | | | | | 1 | 15 | 10 | PET | 43,0 | 17 | 26 |
| PMMA | | | | | | | | | 1 | 1 | PMMA | 2,0 | 0 | 2 |
| | | | | | | | | | | | Polyacetal or POM | 0,0 | 0 | 0 |
| PP | | 1 | 1 | | | | | 4 | 17 | 35 | PP | 58,0 | 2 | 56 |
| PS | 2 | 1 | | | | 1 | 2 | 7 | 41 | 38 | PS | 92,0 | 3 | 89 |
| PU | | | | | | | | | 1 | 2 | PU | 3,0 | 0 | 3 |
| PVC | 1 | | | | | | | 5 | 2 | 3 | PVC | 11,0 | 1 | 10 |
| Tyre | | | | | | | | | | | Tyre | 0,0 | 0 | 0 |
| ABS | | | | | | | | | 2 | 4 | ABS | 6,0 | 0 | 6 |
| rubber | | | | | | | | | 1 | 2 | rubber | 3,0 | 0 | 3 |
| PTFE | | | | | | | | | | | PTFE | 0,0 | 0 | 0 |
| | | | | | | | | | | | | 0,0 | 0 | 0 |
| | | | | | | | | | | | | 0,0 | 0 | 0 |
| Suma | 3,0 | 8,0 | 14,0 | 0,0 | 0,0 | 1,0 | 2,0 | 21,0 | 86,0 | 126,0 | | 261,0 | 25 | 236 |

460

461

462 **Figure SM4. Sample 2 (Subsampling of 4 slices, mean ± standard deviation)**

463

| Polymer | FIBRE | | | | | FRAGMENT | | | | | Polymer | TOTAL | FIBRES | PARTICLES |
|---|---|---|---|---|---|---|---|---|---|---|---|---|---|---|
| | 5000-1000 | 1000-500 | 500-100 | 100-50 | 50-20 | 5000-1000 | 1000-500 | 500-100 | 100-50 | 50-20 | | | | |
| Acrylate | | | | | | | | | | | Acrylate | | | |
| Alkyd Varnish | | | | | | | | | | | Alkyd Varnish | | | |
| EVA | | | | | | | | | | | EVA | | | |
| PA | | | 2,5 ± 5 | | | | | | 2,5 ± 5 | 20 ± 8,1 | PA | 25 ± 10 | 2,5 ± 5 | 22,5 ± 9,5 |
| PE | | 5 ± 10 | | | | | | 2,5 ± 5 | 2,5 ± 5 | | PE | 12,5 ± 18,9 | 5 ± 10 | 7,5 ± 9,5 |
| PC | | | | | | | | | | | PC | | | |
| PET | | | | | | | | | 12,5 ± 5 | 7,5 ± 5 | PET | 20 ± 8,1 | | 20 ± 8,1 |
| PMMA | | | | | | | | | 2,5 ± 5 | 2,5 ± 5 | PMMA | 5 ± 10 | | 5 ± 10 |
| | | | | | | | | | | | Polyacetal or POM | | | |
| PP | | | | | | | | 5 ± 5,7 | 10 ± 8,1 | 42,5 ± 29,8 | PP | 57,5 ± 27,5 | | 57,5 ± 27,5 |
| PS | 2,5 ± 5 | | | | | | | 2,5 ± 5 | 35 ± 12,9 | 47,5 ± 17 | PS | 87,5 ± 26,2 | 2,5 ± 5 | 85 ± 23,8 |
| PU | | | | | | | | | | 2,5 ± 5 | PU | 2,5 ± 5 | | 2,5 ± 5 |
| PVC | | | | | | | | | 2,5 ± 5 | 2,5 ± 5 | PVC | 5 ± 5,7 | | 5 ± 5,7 |
| Tyre | | | | | | | | | | | Tyre | | | |
| ABS | | | | | | | | | 2,5 ± 5 | 2,5 ± 5 | ABS | 5 ± 10 | | 5 ± 10 |
| rubber | | | | | | | | | 2,5 ± 5 | 2,5 ± 5 | rubber | 5 ± 10 | | 5 ± 10 |
| PTFE | | | | | | | | | | | PTFE | | | |
| Suma | 2,5 ± 5 | 5 ± 10 | 2,5 ± 5 | | | | | 10 ± 8,1 | 72,5 ± 22,1 | 132,5 ± 55,6 | | 225 ± 99,4 | 10 ± 20 | 215 ± 83,4 |

464

465

466

467

**Figure SM5. Sample 3 (Entire characterization)**

469

| Polymer | FIBRE | | | | | FRAGMENT | | | | | Polymer | TOTAL | FIBRES | PARTICLES |
|---|---|---|---|---|---|---|---|---|---|---|---|---|---|---|
| | 5000-1000 | 1000-500 | 500-100 | 100-50 | 50-20 | 5000-1000 | 1000-500 | 500-100 | 100-50 | 50-20 | | | | |
| Acrylate | | | | | | | | | | | Acrylate | 0,0 | 0 | 0 |
| Alkyd Varnish | | | | | | | | | | 3 | Alkyd Varnish | 3,0 | 0 | 3 |
| EVA | | | | | | | | | | | EVA | 0,0 | 0 | 0 |
| PA | | | | | | | | | 3 | 141 | PA | 144,0 | 0 | 144 |
| PE | | | 1 | | | | | 1 | 7 | 13 | PE | 22,0 | 1 | 21 |
| PC | | | | | | | | | | | PC | 0,0 | 0 | 0 |
| PET | | 8 | 16 | 1 | | | | 4 | 18 | 57 | PET | 104,0 | 25 | 79 |
| PMMA | | | | | | | | | | | PMMA | 0,0 | 0 | 0 |
| POM | | | | | | | | | | 1 | Polyacetal or POM | 1,0 | 0 | 1 |
| PP | 4 | 3 | 11 | | | | | 13 | 82 | 229 | PP | 342,0 | 18 | 324 |
| PS | | | 1 | | | | | 1 | 4 | 40 | PS | 46,0 | 1 | 45 |
| PU | | | | | | | | | | 3 | PU | 3,0 | 0 | 3 |
| PVC | | | | | | | | | | 1 | PVC | 1,0 | 0 | 1 |
| Tyre | | | | | | | | | | | Tyre | 0,0 | 0 | 0 |
| ABS | | | 1 | | | | | 1 | 6 | 46 | ABS | 54,0 | 1 | 53 |
| rubber | | | | | | | | | | 6 | rubber | 6,0 | 0 | 6 |
| PTFE | | | | | | | | | | | PTFE | 0,0 | 0 | 0 |
| | | | | | | | | | | | | 0,0 | 0 | 0 |
| | | | | | | | | | | | | 0,0 | 0 | 0 |
| Suma | 4,0 | 11,0 | 30,0 | 1,0 | 0,0 | 0,0 | 0,0 | 20,0 | 120,0 | 540,0 | | 726,0 | 46 | 680 |

470

471

**Figure SM6. Sample 3 (Subsampling of 4 slices, mean ± standard deviation)**

| Polymer | FIBRE | | | | | FRAGMENT | | | | | Polymer | TOTAL | FIBRES | PARTICLES |
|---|---|---|---|---|---|---|---|---|---|---|---|---|---|---|
| | 5000-1000 | 1000-500 | 500-100 | 100-50 | 50-20 | 5000-1000 | 1000-500 | 500-100 | 100-50 | 50-20 | | | | |
| Acrylate | | | | | | | | | | | Acrylate | | | |
| Alkyd Varnish | | | | | | | | | | 2,5 ± 5 | Alkyd Varnish | 2,5 ± 5 | | 2,5 ± 5 |
| EVA | | | | | | | | | | | EVA | | | |
| PA | | | | | | | | | 2,5 ± 5 | 142,5 ± 100,1 | PA | 145 ± 98,1 | | 145 ± 98,1 |
| PE | | | 2,5 ± 5 | | | | | | 2,5 ± 5 | 7,5 ± 9,5 | PE | 12,5 ± 15 | 2,5 ± 5 | 10 ± 14,1 |
| PC | | | | | | | | | | | PC | | | |
| PET | | 10 ± 8,1 | 25 ± 12,9 | 2,5 ± 5 | | | | | 15 ± 12,9 | 70 ± 18,2 | PET | 122,5 ± 33 | 37,5 ± 17 | 85 ± 25,1 |
| PMMA | | | | | | | | | | | PMMA | | | |
| POM | | | | | | | | | | | Polyacetal or POM | | | |
| PP | 5 ± 5,7 | | 10 ± 8,1 | | | | | 2,5 ± 5 | 82,5 ± 45 | 252,5 ± 137,2 | PP | 352,5 ± 180,8 | 15 ± 12,9 | 337,5 ± 169,7 |
| PS | | 2,5 ± 5 | | | | | | | 5 ± 10 | 52,5 ± 42,7 | PS | 60 ± 43,2 | 2,5 ± 5 | 57,5 ± 38,6 |
| PU | | | | | | | | | | 2,5 ± 5 | PU | 2,5 ± 5 | | 2,5 ± 5 |
| PVC | | | | | | | | | | | PVC | | | |
| Tyre | | | | | | | | | | | Tyre | | | |
| ABS | | | | | | | | | 10 ± 14,1 | 47,5 ± 61,8 | ABS | 57,5 ± 75 | | 57,5 ± 75 |
| rubber | | | | | | | | | | 5 ± 5,7 | rubber | 5 ± 5,7 | | 5 ± 5,7 |
| PTFE | | | | | | | | | | | PTFE | | | |
| Suma | 5 ± 5,7 | 10 ± 8,1 | 40 ± 24,4 | 2,5 ± 5 | | | | 2,5 ± 5 | 117,5 ± 63,9 | 582,5 ± 241,7 | | 760 ± 298,7 | 57,5 ± 32 | 702,5 ± 278,2 |

473

474

475

476

477

478

479 **Figure SM7. Sample 4 (Entire characterization)**

| Polymer | FIBRE 5000-1000 | 1000-500 | 500-100 | 100-50 | 50-20 | FRAGMENT 5000-1000 | 1000-500 | 500-100 | 100-50 | 50-20 | Polymer | TOTAL | FIBRES | PARTICLES |
|---|---|---|---|---|---|---|---|---|---|---|---|---|---|---|
| Acrylate | | | 1 | | | | | 1 | 1 | 1 | Acrylate | 4,0 | 1 | 3 |
| Alkyd Varnish | | | | | | | | | | | Alkyd Varnish | 0,0 | 0 | 0 |
| EVA | | | | | | | | | | | EVA | 0,0 | 0 | 0 |
| PA | | | | | | | | | 3 | 22 | PA | 25,0 | 0 | 25 |
| PE | | | 1 | | | | | 3 | 8 | 21 | PE | 33,0 | 1 | 32 |
| PC | | | | | | | | | | | PC | 0,0 | 0 | 0 |
| PET | 1 | 5 | 19 | | | | | 15 | 22 | 42 | PET | 104,0 | 25 | 79 |
| PMMA | | | | | | | | | | | PMMA | 0,0 | 0 | 0 |
| POM | | | | | | | | | | | Polyacetal or POM | 0,0 | 0 | 0 |
| PP | 1 | | 5 | 2 | | 1 | 3 | 13 | 54 | 235 | PP | 314,0 | 8 | 306 |
| PS | | | | | | | | | 2 | 11 | PS | 13,0 | 0 | 13 |
| PU | | | | | | | | | | 2 | PU | 2,0 | 0 | 2 |
| PVC | | | | | | | | | 1 | 2 | PVC | 3,0 | 0 | 3 |
| Tyre | | | | | | | | | | | Tyre | 0,0 | 0 | 0 |
| ABS | | | | | | | | | 4 | 7 | ABS | 11,0 | 0 | 11 |
| rubber | | | | | | | | | | 12 | rubber | 12,0 | 0 | 12 |
| PTFE | | | | | | | | | 1 | | PTFE | 1,0 | 0 | 1 |
| | | | | | | | | | | | | 0,0 | 0 | 0 |
| | | | | | | | | | | | | 0,0 | 0 | 0 |
| Suma | 2,0 | 5,0 | 26,0 | 2,0 | 0,0 | 1,0 | 3,0 | 32,0 | 96,0 | 355,0 | | 522,0 | 35 | 487 |

480

481

482 **Figure SM8. Sample 4 (Subsampling of 4 slices, mean ± standard deviation)**

| Polymer | FIBRE 5000-1000 | 1000-500 | 500-100 | 100-50 | 50-20 | FRAGMENT 5000-1000 | 1000-500 | 500-100 | 100-50 | 50-20 | Polymer | TOTAL | FIBRES | PARTICLES |
|---|---|---|---|---|---|---|---|---|---|---|---|---|---|---|
| Acrylate | | | | | | | | 2,5 ± 5 | | | Acrylate | 2,5 ± 5 | | 2,5 ± 5 |
| Alkyd Varnish | | | | | | | | | | | Alkyd Varnish | | | |
| EVA | | | | | | | | | | | EVA | | | |
| PA | | | | | | | | | 2,5 ± 5 | 10 ± 14,1 | PA | 12,5 ± 15 | | 12,5 ± 15 |
| PE | | | | | | | | | 5 ± 10 | 32,5 ± 26,2 | PE | 37,5 ± 22,1 | | 37,5 ± 22,1 |
| PC | | | | | | | | | | | PC | | | |
| PET | 2,5 ± 5 | 10 ± 8,1 | 17,5 ± 20,6 | | | | | 2,5 ± 5 | 17,5 ± 9,5 | 40 ± 14,1 | PET | 90 ± 20 | 30 ± 29,4 | 60 ± 27 |
| PMMA | | | | | | | | | | | PMMA | | | |
| POM | | | | | | | | | | | Polyacetal or POM | | | |
| PP | 2,5 ± 5 | | 2,5 ± 5 | 2,5 ± 5 | | 2,5 ± 5 | 2,5 ± 5 | 7,5 ± 15 | 45 ± 17,3 | 227,5 ± 18,9 | PP | 292,5 ± 43,4 | 7,5 ± 5 | 285 ± 45 |
| PS | | | | | | | | | | 17,5 ± 23,6 | PS | 17,5 ± 23,6 | | 17,5 ± 23,6 |
| PU | | | | | | | | | | | PU | | | |
| PVC | | | | | | | | | | 2,5 ± 5 | PVC | 2,5 ± 5 | | 2,5 ± 5 |
| Tyre | | | | | | | | | | | Tyre | | | |
| ABS | | | | | | | | | 2,5 ± 5 | 15 ± 5,7 | ABS | 17,5 ± 5 | | 17,5 ± 5 |
| rubber | | | | | | | | | 2,5 ± 5 | 12,5 ± 18,9 | rubber | 15 ± 23,8 | | 15 ± 23,8 |
| PTFE | | | | | | | | | | | PTFE | | | |
| Suma | 5 ± 10 | 10 ± 8,1 | 20 ± 18,2 | 2,5 ± 5 | | 2,5 ± 5 | 2,5 ± 5 | 12,5 ± 15 | 75 ± 23,8 | 357,5 ± 62,9 | | 487,5 ± 35 | 37,5 ± 27,5 | 450 ± 57,1 |

483

484

485

486

**Figure SM9. Sample 5 (Blank 1) (Entire characterization)**

| Polymer | FIBRE 5000-1000 | 1000-500 | 500-100 | 100-50 | 50-20 | FRAGMENT 5000-1000 | 1000-500 | 500-100 | 100-50 | 50-20 | Polymer | TOTAL | FIBRES | PARTICLES |
|---|---|---|---|---|---|---|---|---|---|---|---|---|---|---|
| Acrylate | | | | | | | | | | | Acrylate | 0,0 | 0 | 0 |
| Alkyd Varnish | | | | | | | | | | | Alkyd Varnish | 0,0 | 0 | 0 |
| EVA | | | | | | | | | | | EVA | 0,0 | 0 | 0 |
| PA | | | | | | | | | | | PA | 0,0 | 0 | 0 |
| PE | | | | | | | | | | | PE | 0,0 | 0 | 0 |
| PC | | | | | | | | | | | PC | 0,0 | 0 | 0 |
| PET | | | | | | | | 1 | | | PET | 1,0 | 0 | 1 |
| PMMA | | | | | | | | | | | PMMA | 0,0 | 0 | 0 |
| POM | 1 | | 3 | 1 | | | | | | | Polyacetal or POM | 0,0 | 0 | 0 |
| PP | | | | | | | | 1 | 2 | 15 | PP | 23,0 | 5 | 18 |
| PS | | | | | | | | | | 1 | PS | 1,0 | 0 | 1 |
| PU | | | | | | | | | | 1 | PU | 1,0 | 0 | 1 |
| PVC | | | | | | | | | | | PVC | 0,0 | 0 | 0 |
| Tyre | | | | | | | | 1 | 2 | 9 | Tyre | 12,0 | 0 | 12 |
| ABS | | | | | | | | | | | ABS | 0,0 | 0 | 0 |
| rubber | | | | | | | | 3 | 4 | 5 | rubber | 12,0 | 0 | 12 |
| PTFE | | | | | | | | | | | PTFE | 0,0 | 0 | 0 |
| | | | | | | | | | | | | 0,0 | 0 | 0 |
| | | | | | | | | | | | | 0,0 | 0 | 0 |
| Suma | 1,0 | 0,0 | 3,0 | 1,0 | 0,0 | 0,0 | 0,0 | 6,0 | 8,0 | 31,0 | | 50,0 | 5 | 45 |

488

489

**Figure SM10. Blank 1 (Subsampling of 4 slices, mean ± standard deviation)**

0

| Polymer | FIBRE 5000-1000 | 1000-500 | 500-100 | 100-50 | 50-20 | FRAGMENT 5000-1000 | 1000-500 | 500-100 | 100-50 | 50-20 | Polymer | TOTAL | FIBRES | PARTICLES |
|---|---|---|---|---|---|---|---|---|---|---|---|---|---|---|
| Acrylate | | | | | | | | | | | Acrylate | | | |
| Alkyd Varnish | | | | | | | | | | | Alkyd Varnish | | | |
| EVA | | | | | | | | | | | EVA | | | |
| PA | | | | | | | | | | | PA | | | |
| PE | | | | | | | | | | | PE | | | |
| PC | | | | | | | | | | | PC | | | |
| PET | | | | | | | | 2,5 ± 5 | | | PET | 2,5 ± 5 | | 2,5 ± 5 |
| PMMA | | | | | | | | | | | PMMA | | | |
| POM | | | | | | | | | | | Polyacetal or POM | | | |
| PP | 2,5 ± 5 | | 2,5 ± 5 | | | | | 2,5 ± 5 | 5 ± 5,7 | 15 ± 23,8 | PP | 27,5 ± 27,5 | 5 ± 10 | 22,5 ± 26,2 |
| PS | | | | | | | | | | | PS | | | |
| PU | | | | | | | | | | 2,5 ± 5 | PU | 2,5 ± 5 | | 2,5 ± 5 |
| PVC | | | | | | | | | | | PVC | | | |
| Tyre | | | | | | | | | 2,5 ± 5 | 5 ± 10 | Tyre | 7,5 ± 15 | | 7,5 ± 15 |
| ABS | | | | | | | | | | | ABS | | | |
| rubber | | | | | | | | 5 ± 10 | 2,5 ± 5 | 5 ± 5,7 | rubber | 12,5 ± 9,5 | | 12,5 ± 9,5 |
| PTFE | | | | | | | | | | | PTFE | | | |
| Suma | 2,5 ± 5 | | 2,5 ± 5 | | | | | 10 ± 8,1 | 10 ± 0 | 27,5 ± 17 | | 52,5 ± 22,1 | 5 ± 10 | 47,5 ± 22,1 |

493

494

495

496

497

**Figure SM11. Sample 6. Blank 2 (Entire characterization)**

499

| Polymer | FIBRE 5000-1000 | 1000-500 | 500-100 | 100-50 | 50-20 | FRAGMENT 5000-1000 | 1000-500 | 500-100 | 100-50 | 50-20 | Polymer | TOTAL | FIBRES | PARTICLES |
|---|---|---|---|---|---|---|---|---|---|---|---|---|---|---|
| Acrylate | | | | | | | | | | | Acrylate | 0,0 | 0 | 0 |
| Alkyd Varnish | | | | | | | | | | | Alkyd Varnish | 0,0 | 0 | 0 |
| EVA | | | | | | | | | | | EVA | 0,0 | 0 | 0 |
| PA | | | | | | | | | | 8 | PA | 8,0 | 0 | 8 |
| PE | | | | | | | | | | | PE | 0,0 | 0 | 0 |
| PC | | | | | | | | | | | PC | 0,0 | 0 | 0 |
| PET | | | 1 | | | | | 1 | 3 | 1 | PET | 6,0 | 1 | 5 |
| PMMA | | | | | | | | | | | PMMA | 0,0 | 0 | 0 |
| POM | | | | | | | | | | 1 | Polyacetal or POM | 1,0 | 0 | 1 |
| PP | | | | | | | | | 1 | 1 | PP | 2,0 | 0 | 2 |
| PS | | | | | | | | | | | PS | 0,0 | 0 | 0 |
| PU | | | | | | | | | | | PU | 0,0 | 0 | 0 |
| PVC | | | | | | | | | | | PVC | 0,0 | 0 | 0 |
| Tyre | | | | | | | | | | | Tyre | 0,0 | 0 | 0 |
| ABS | | | | | | | | | | 7 | ABS | 7,0 | 0 | 7 |
| rubber | | | | | | | | | 2 | 7 | rubber | 9,0 | 0 | 9 |
| PTFE | | | | | | | | | | | PTFE | 0,0 | 0 | 0 |
| | | | | | | | | | | | | 0,0 | 0 | 0 |
| | | | | | | | | | | | | 0,0 | 0 | 0 |
| Suma | 0,0 | 0,0 | 1,0 | 0,0 | 0,0 | 0,0 | 0,0 | 1,0 | 6,0 | 25,0 | | 33,0 | 1 | 32 |

**Figure SM12. Blank 2 (Subsampling of 4 slices, mean ± standard deviation)**

| Polymer | FIBRE 5000-1000 | 1000-500 | 500-100 | 100-50 | 50-20 | FRAGMENT 5000-1000 | 1000-500 | 500-100 | 100-50 | 50-20 | Polymer | TOTAL | FIBRES | PARTICLES |
|---|---|---|---|---|---|---|---|---|---|---|---|---|---|---|
| Acrylate | | | | | | | | | | | Acrylate | | | |
| Alkyd Varnish | | | | | | | | | | | Alkyd Varnish | | | |
| EVA | | | | | | | | | | | EVA | | | |
| PA | | | | | | | | | | 12,5 ± 12,5 | PA | 12,5 ± 12,5 | | 12,5 ± 12,5 |
| PE | | | | | | | | | | | PE | | | |
| PC | | | | | | | | | | | PC | | | |
| PET | | | | | | | | | | 2,5 ± 5 | PET | 2,5 ± 5 | | 2,5 ± 5 |
| PMMA | | | | | | | | | | | PMMA | | | |
| POM | | | | | | | | | | | Polyacetal or POM | | | |
| PP | | | | | | | | | 2,5 ± 5 | | PP | 2,5 ± 5 | | 2,5 ± 5 |
| PS | | | | | | | | | | | PS | | | |
| PU | | | | | | | | | | | PU | | | |
| PVC | | | | | | | | | | | PVC | | | |
| Tyre | | | | | | | | | | | Tyre | | | |
| ABS | | | | | | | | | | 15 ± 17,3 | ABS | 15 ± 17,3 | | 15 ± 17,3 |
| rubber | | | | | | | | | | 7,5 ± 9,5 | rubber | 7,5 ± 9,5 | | 7,5 ± 9,5 |
| PTFE | | | | | | | | | | | PTFE | | | |
| Suma | | | | | | | | | 2,5 ± 5 | 37,5 ± 17 | | 40 ± 16,3 | | 40 ± 16,3 |

502

503

504

505

506

**Figure SM13. Sample 5 (Gold-coated filter, Entire characterization)**

| Polymer | FIBRE | | | | | FRAGMENT | | | | | Polymer | TOTAL | FIBRES | PARTICLES |
|---|---|---|---|---|---|---|---|---|---|---|---|---|---|---|
| | 5000-1000 | 1000-500 | 500-100 | 100-50 | 50-20 | 5000-1000 | 1000-500 | 500-100 | 100-50 | 50-20 | | | | |
| Acrylate | | | | | | | | | 1 | 2 | Acrylate | 3,0 | 0 | 3 |
| Alkyd Varnish | | | | | | | | | | | Alkyd Varnish | 0,0 | 0 | 0 |
| EVA | | 1 | | | | | | | | | EVA | 1,0 | 1 | 0 |
| PA | | | | | | | | | | 10 | PA | 10,0 | 0 | 10 |
| PE | | 2 | 2 | 2 | | | | 1 | 13 | 21 | PE | 41,0 | 6 | 35 |
| PC | | | | | | | | | 1 | 1 | PC | 2,0 | 0 | 2 |
| PET | | | 2 | | | | | | 3 | 7 | PET | 12,0 | 2 | 10 |
| PMMA | | | | | | | | | | 1 | PMMA | 1,0 | 0 | 1 |
| POM | | | | | | | | | | | Polyacetal or POM | 0,0 | 0 | 0 |
| PP | | 4 | 12 | 2 | | | | 5 | 30 | 120 | PP | 173,0 | 18 | 155 |
| PS | | | 1 | 1 | | | | | 1 | 7 | PS | 10,0 | 2 | 8 |
| PU | 1 | 1 | | | | | | 1 | 2 | 5 | PU | 10,0 | 2 | 8 |
| PVC | | | | | | | | | | 1 | PVC | 0,0 | 0 | 1 |
| Tyre | | | | | | | | | | | Tyre | 0,0 | 0 | 0 |
| ABS | | | | | | | | | | 6 | ABS | 6,0 | 0 | 6 |
| rubber | | | | | | | | | | | rubber | 0,0 | 0 | 0 |
| PTFE | | | | | | | | | | | PTFE | 0,0 | 0 | 0 |
| | | | | | | | | | | | | 0,0 | 0 | 0 |
| | | | | | | | | | | | | 0,0 | 0 | 0 |
| Suma | 1,0 | 8,0 | 17,0 | 5,0 | 0,0 | 0,0 | 0,0 | 7,0 | 51,0 | 181,0 | | 270,0 | 31 | 239 |

508

509

**Figure SM14. Sample 5 (Gold-coated filter, subsampling of 4 slices, mean ± standard deviation)**

| Polymer | FIBRE | | | | | FRAGMEMT | | | | | Polymer | TOTAL | FIBRES | PARTICLES |
|---|---|---|---|---|---|---|---|---|---|---|---|---|---|---|
| | 5000-1000 | 1000-500 | 500-100 | 100-50 | 50-20 | 5000-1000 | 1000-500 | 500-100 | 100-50 | 50-20 | | | | |
| Acrylate | | | | | | | | | | | Acrylate | | | |
| Alkyd Varnish | | | | | | | | | | | Alkyd Varnish | | | |
| EVA | | 2,5 ± 5 | | | | | | | | | EVA | 2,5 ± 5 | 2,5 ± 5 | |
| PA | | | | | | | | | | 9,2 ± 10,7 | PA | 9,2 ± 10,7 | | 9,2 ± 10,7 |
| PE | | | | 2,1 ± 4,2 | | | | 6,1 ± 12,2 | 4,6 ± 5,3 | 25,9 ± 11,4 | PE | 38,8 ± 18,9 | 2,1 ± 4,2 | 36,7 ± 16,1 |
| PC | | | | | | | | | | | PC | | | |
| PET | | | 2,1 ± 4,2 | | | | | | 8,2 ± 11,5 | 10,7 ± 10,1 | PET | 21 ± 18,8 | 2,1 ± 4,2 | 18,9 ± 19,9 |
| PMMA | | | | | | | | | | | PMMA | | | |
| POM | | | | | | | | | | | Polyacetal or POM | | | |
| PP | | 2,1 ± 4,2 | 16,3 ± 17,2 | | | | | 10,3 ± 10,1 | 39,9 ± 17,3 | 119,6 ± 12,2 | PP | 188,4 ± 20,3 | 18,5 ± 16,4 | 169,9 ± 20,6 |
| PS | | | | 2,1 ± 4,2 | | | | | | 12,4 ± 14,4 | PS | 14,6 ± 12,4 | 2,1 ± 4,2 | 12,4 ± 14,4 |
| PU | | | | | | | | | 2,1 ± 4,2 | 7,1 ± 9,4 | PU | 9,2 ± 8,2 | | 9,2 ± 8,2 |
| PVC | | | | | | | | | | | PVC | | | |
| Tyre | | | | | | | | | | | Tyre | | | |
| ABS | | | | | | | | | | 4,6 ± 5,3 | ABS | 4,6 ± 5,3 | | 4,6 ± 5,3 |
| rubber | | | | | | | | | | | rubber | | | |
| PTFE | | | | | | | | | | | PTFE | | | |
| Suma | | 4,6 ± 5,3 | 18,5 ± 17,8 | 4,2 ± 4,9 | | | | 16,4 ± 21,9 | 54,9 ± 19,6 | 189,8 ± 14,4 | | 288,5 ± 32,2 | 27,3 ± 20,8 | 261,1 ± 43 |

512

513

514

515 **Figure SM15. Sample 6 (Gold-coated filter, Entire characterization)**

| Polymer | FIBRE 5000-1000 | 1000-500 | 500-100 | 100-50 | 50-20 | FRAGMENT 5000-1000 | 1000-500 | 500-100 | 100-50 | 50-20 | Polymer | TOTAL | FIBRES | PARTICLES |
|---|---|---|---|---|---|---|---|---|---|---|---|---|---|---|
| Acrylate | 1 | | | | | | | 1 | 1 | | Acrylate | 3,0 | 1 | 2 |
| Alkyd Varnish | | | | | | | | | | | Alkyd Varnish | 0,0 | 0 | 0 |
| EVA | | | | | | | | | | 1 | EVA | 1,0 | 0 | 1 |
| PA | | | 1 | | | | | | 1 | 3 | PA | 5,0 | 1 | 4 |
| PE | | | 15 | 2 | | | | 14 | 94 | 186 | PE | 311,0 | 17 | 294 |
| PC | | | | | | | | | | | PC | 0,0 | 0 | 0 |
| PET | | 2 | 5 | | | | | 1 | | 4 | PET | 12,0 | 7 | 5 |
| PMMA | | | | | | | | | | 2 | PMMA | 2,0 | 0 | 2 |
| POM | | | | | | | | | | | Polyacetal or POM | 0,0 | 0 | 0 |
| PP | | 1 | 8 | 4 | | | | 1 | 4 | 39 | PP | 57,0 | 13 | 44 |
| PS | | | | | | | | | | 1 | PS | 1,0 | 0 | 1 |
| PU | | | | | | | | | 1 | 6 | PU | 7,0 | 0 | 7 |
| PVC | | | | | | | | | 1 | | PVC | 1,0 | 0 | 1 |
| Tyre | | | | | | | | | | | Tyre | 0,0 | 0 | 0 |
| ABS | | | | | | | | | 1 | 2 | ABS | 3,0 | 0 | 3 |
| rubber | | | | | | | | | | | rubber | 0,0 | 0 | 0 |
| PTFE | | | | | | | | | | | PTFE | 0,0 | 0 | 0 |
| | | | | | | | | | | | | 0,0 | 0 | 0 |
| | | | | | | | | | | | | 0,0 | 0 | 0 |
| Suma | 1,0 | 3,0 | 29,0 | 6,0 | 0,0 | 0,0 | 0,0 | 17,0 | 103,0 | 244,0 | | 403,0 | 39 | 364 |

516

517

518 **Figure SM16. Sample 6 (Gold-coated filter, subsampling of 4 slices, mean ± standard**
519 **deviation)**

520

| Polymer | FIBRE 5000-1000 | 1000-500 | 500-100 | 100-50 | 50-20 | FRAGMENT 5000-1000 | 1000-500 | 500-100 | 100-50 | 50-20 | Polymer | TOTAL | FIBRES | PARTICLES |
|---|---|---|---|---|---|---|---|---|---|---|---|---|---|---|
| Acrylate | 2,1 ± 4,2 | | | | | | | | | | Acrylate | 2,1 ± 4,2 | 2,1 ± 4,2 | |
| Alkyd Varnish | | | | | | | | | | | Alkyd Varnish | | | |
| EVA | | | | | | | | | | | EVA | | | |
| PA | | | 2,1 ± 4,2 | | | | | | | 8,2 ± 11,4 | PA | 10,3 ± 12,3 | 2,1 ± 4,2 | 8,2 ± 11,4 |
| PE | | | 7,1 ± 9,4 | | | | | 8,5 ± 9,8 | 98 ± 36,8 | 220,1 ± 49,6 | PE | 333,8 ± 45 | 7,1 ± 9,4 | 326,6 ± 45,9 |
| PC | | | | | | | | | | | PC | | | |
| PET | | | 17,1 ± 23 | | | | | 2,1 ± 4,2 | | 2,1 ± 4,2 | PET | 21,4 ± 20,2 | 17,1 ± 23 | 4,2 ± 8,5 |
| PMMA | | | | | | | | | | 2,1 ± 4,2 | PMMA | 2,1 ± 4,2 | | 2,1 ± 4,2 |
| POM | | | | | | | | | | | Polyacetal or POM | | | |
| PP | | | 6,7 ± 4,5 | 6,7 ± 8,3 | | | | 6 ± 12,1 | 2,1 ± 4,2 | 42 ± 26,7 | PP | 63,7 ± 29,4 | 13,5 ± 11,4 | 50,2 ± 36,1 |
| PS | | | | | | | | | | 2,1 ± 4,2 | PS | 2,1 ± 4,2 | | 2,1 ± 4,2 |
| PU | | | | | | | | | 6 ± 12,1 | 9,2 ± 8,2 | PU | 15,3 ± 8,1 | | 15,3 ± 8,1 |
| PVC | | | | | | | | | | | PVC | | | |
| Tyre | | | | | | | | | | | Tyre | | | |
| ABS | | | | | | | | | | 2,1 ± 4,2 | ABS | 2,1 ± 4,2 | | 2,1 ± 4,2 |
| rubber | | | | | | | | | | | rubber | | | |
| PTFE | | | | | | | | | | | PTFE | | | |
| Suma | 2,1 ± 4,2 | | 33,1 ± 18,7 | 6,7 ± 8,3 | | | | 16,7 ± 11,7 | 106,2 ± 36,1 | 288,1 ± 74,9 | | 453,2 ± 89 | 42 ± 15,3 | 411,1 ± 84,1 |

521

522

523

524

### Sample 5

| Polymer | FIBRE 5000-1000 | 1000-500 | 500-100 | 100-50 | 50-20 | FRAGMENT 5000-1000 | 1000-500 | 500-100 | 100-50 | 50-20 |
|---|---|---|---|---|---|---|---|---|---|---|
| Acrylate | | | | | | | | | 0.0 | 0.0 |
| Alkyd Varnish | | | | | | | | | | |
| EVA | | 250.0 | | | | | | | | |
| PA | | | | | | | | | | 92.5 |
| PE | | 0.0 | 0.0 | 106.3 | | | | 610.0 | 35.6 | 123.7 |
| PC | | | | | | | | | 0.0 | 0.0 |
| PET | | | 106.3 | | | | | | 274.2 | 153.2 |
| PMMA | | | | | | | | | | 0.0 |
| Polyacetal or POM | | | | | | | | | | |
| PP | | 53.1 | 136.5 | 0.0 | | | | 207.0 | 133.2 | 99.7 |
| PS | | | 0.0 | 212.5 | | | | | 0.0 | 178 |
| PU | 0.0 | | | | | | | 0.0 | 106.3 | 142.5 |
| PVC | | | | | | | | | | 0.0 |
| Tyre | | | | | | | | | | |
| ABS | | | | | | | | | | 77.1 |
| rubber | | | | | | | | | | |
| PTFE | | | | | | | | | | |

### Sample 6

| Polymer | FIBRE 5000-1000 | 1000-500 | 500-100 | 100-50 | 50-20 | FRAGMENT 5000-1000 | 1000-500 | 500-100 | 100-50 | 50-20 |
|---|---|---|---|---|---|---|---|---|---|---|
| Acrylate | 212.5 | | | | | | | 0.0 | 0.0 | |
| Alkyd Varnish | | | | | | | | | | |
| EVA | | | | | | | | | | 0.0 |
| PA | | | 212.5 | | | | | | 0.0 | 274.1 |
| PE | | | 47.5 | 0.0 | | | | 60.7 | 104.3 | 118.4 |
| PC | | | | | | | | | | |
| PET | | 0.0 | 343.9 | | | | | 212.5 | | 53.1 |
| PMMA | | | | | | | | | | 106.3 |
| Polyacetal or POM | | | | | | | | | | |
| PP | | 0.0 | 84.4 | 168.8 | | | | 609.8 | 53.1 | 107.8 |
| PS | | | | | | | | | | 212.5 |
| PU | | | | | | | | | 609.8 | 154.2 |
| PVC | | | | | | | | | 0.0 | |
| Tyre | | | | | | | | | | |
| ABS | | | | | | | | 0.0 | | 106.3 |
| rubber | | | | | | | | | | |
| PTFE | | | | | | | | | | |

525

526 **Figure SM17**. Recoveries obtained for the sample-based subsampling strategy when applied to a
527 gold-coated filter, as a function of polymer and size. Blue =statistical coincidence with reference
528 value (68 % confidence level), red = no identification when subsampling.

529